\documentclass[aps,prb,twocolumn,showpacs,superscriptaddress,groupedaddress]{revtex4}  % for review and submission

\usepackage[version=3]{mhchem} % Formula subscripts using \ce{}
\usepackage{times}
\usepackage{graphicx}
\usepackage[caption=false]{subfig}

\newcommand{\unitvec}[1]{\hat{\vec{#1}}}
\renewcommand{\vec}[1]{\mathbf{#1}}

\begin{document}

\title{Design of Tunable Biperiodic Graphene Metasurfaces}

\author{Arya Fallahi} \affiliation{DESY-Center for Free-Electron Laser Science, Hamburg University, Notkestrasse 85, D-22607 Hamburg, Germany}
\author{Julien Perruisseau-Carrier} \affiliation{Ecole Polytechnique F\'{e}d\'{e}rale de Lausanne (EPFL), CH-1015 Lausanne, Switzerland}

\date{\today}

\begin{abstract}
Periodic structures with subwavelength features are instrumental in the versatile and effective control of electromagnetic waves from radio frequencies up to optics.
In this paper, we theoretically evaluate the potential applications and performance of electromagnetic metasurfaces made of periodically patterned graphene. %%
Several graphene metasurfaces are presented, thereby demonstrating that such ultrathin surfaces can be used to dynamically control the electromagnetic wave reflection, absorption, or polarization. %%
Indeed, owing to the graphene properties, the structure performance in terms of resonance frequencies and bandwidths changes with the variation of electrostatic bias fields. %%
To demonstrate the applicability of the concept at different frequency ranges, the examples provided range from microwave to infrared, corresponding to graphene features with length-scales of a few millimeters down to about a micrometer, respectively. %%
The results are obtained using a full-vector semi-analytical numerical technique developed to accurately model the graphene-based multilayer periodic structures under study.
\end{abstract}

\maketitle


\section{Introduction}

Graphene is a flat monolayer of carbon atoms arranged in a honeycomb lattice, which can also be viewed as the building block for other graphitic materials, such as fullerenes, carbon nanotubes and bulk graphite \cite{theRiseOfGraphene, grapheneStatusAndProspects}. %%
The experiments on graphene confirmed that its electrons behave as massless Dirac fermions with the possibility of traversing long distances without being scattered by the crystal, which in turn results in the capability to sustain large electric currents \cite{masslessDiracFermions1,masslessDiracFermions2,masslessDiracFermions3,masslessDiracFermions4}. %%
These unprecedented properties underpinned extensive applications and development of ultrathin carbon nanoelectronic active devices using small graphene samples \cite{nanoelectronicDevicesGraphene,grapheneTransistors}. %%
In the early years of graphene research, it was one of the most expensive materials obtained through the nanofabrication techniques \cite{grapheneFabrication1}. %%
However, recently efficient growth of large graphene samples has been achieved using chemical vapor deposition, which has now become a common graphene fabrication technique \cite{grapheneFabrication2,grapheneFabrication3}. %%
This allowed the production of larger graphene based structures as needed to realize the graphene electromagnetic (EM) metasurfaces discussed in this paper.%%

One novel area, where the properties of graphene may be influential, is the dynamic control of the EM waves propagation \cite{grapheneMetamaterials1,grapheneMetamaterials2,grapheneMetamaterials3,grapheneMetamaterials4}. %%
Notably, the capability of graphene to serve as a platform in transformation optics has been shown based on a homogeneous (i.e. unpatterned) but spatially bias modulated  graphene layer represented by a free-standing surface in a commercial EM solver \cite{grapheneMetamaterials1,sebasProceedingPaper}.
More recently, employing graphene to realize cloaking surfaces has also been proposed \cite{grapheneMetamaterials4}. %%
Furthermore, employing a graphene monolayer for sensing and measuring a metamaterial performance was also suggested \cite{grapheneMetamaterials2}.
These studies on graphene EM metasurfaces considered a homogeneous layer of graphene. %%
Only very recently, research effort has been devoted on nanostructuring graphene to control its effective properties. %%
For instance, the use of graphene strips for the realization of plasmonic waveguides with confined field profiles have been investigated \cite{grapheneStripes1}.
The possibility of controlling the plasmonic resonances in a periodic arrangement of such strips has also been reported \cite{grapheneStripes2}. %%
The effect of patterning on the transmission properties of this monoatomic layers were studied as well \cite{patternedGrapheneNature}. %%
Finally, the use of periodic graphene metasurfaces to achieve strong absorption of infra-red radiation was reported \cite{grapheneAbsorption1,grapheneAbsorption2,grapheneAbsorption3,grapheneAbsorption4}. %%

The biperiodic patterning of \emph{metal} has been long used at micro and millimeter-wave frequencies to achieve electromagnetic properties which cannot be obtained using uniform multilayered structures. %%
Perhaps, the most widespread application of such periodic screens lies in frequency-selective surfaces (FSS), which are spectral filters for free space propagation applications (as employed in radar technology) \cite{bookFSS1,bookFSS2}. %%
Though different FSSs having fixed frequency responses are successfully modeled, implemented and utilized, their \emph{dynamic} control is still a challenging issue.
The mechanism allowing the dynamic control of the FSS performance generally introduces very tough limitations on the achievable characteristics. %%
In the microwave regime, the use of liquid crystals, varactor diodes, and MEMS switches to provide a dynamic frequency response has been reported \cite{dynamicFSS1,dynamicFSS2,dynamicFSS3,dynamicFSS4}.
However, achieving such characteristics in the terahertz and infrared regime is even more difficult because of the aforementioned technological reasons.
In this context, the use of micro/nano-patterned graphene seems extremely interesting to overcome the limitations of existing technologies, in terms of operation frequency, biasing complexity (thanks to graphene well-known electric field effect \cite{grapheneConductivityModel,grapheneOpticalProperties}), as well as integration and miniaturization. %%

We demonstrate this potential by presenting the modeling and design of graphene-based dynamically-controllable metasurface, at both microwave and THz frequencies.
First, graphene conductivity model is recalled and the numerical algorithm used for modeling the electromagnetic interaction of a plane wave with a patterned graphene metasurface is presented. %%
Subsequently, application examples are analyzed using the proposed method. %%
It is shown that electronic gating of a miniaturized graphene layer allows one to control both the appearance and the position of the periodic structure resonances.
Results for a controllable X-band microwave absorber and an infrared switchable polarizer are presented, highlighting the possibility to also affect absorption and polarization through anisotropic structuring.
The designed structures are based on multilayer graphene stacks, since conventional biasing techniques cannot be employed without shielding the metasurface effective response (note that a similar idea has recently been proposed for fabricating THz modulators \cite{grapheneModulators}). %%
The different examples show that graphene not only enables convenient dynamic control of the structure response, but also that its periodic patterning also allows tailoring the response both in terms of general functionality and operation frequencies.

\section{Graphene Conductivity Model}

A graphene sheet can be modeled as an infinitesimally thin conductive sheet, whose conductivity is obtained using a semi-classical quantum mechanical method. %%
In low frequency regime (low THz regime and below) and in the absence of magnetostatic biasing, this conductivity is represented by a scalar value \cite{grapheneModeling3}. %%
However, at higher THz and mid infra-red frequencies for very intense spatial variations of the interacting wave (in the scale of the scattering length), the conductivity may possess non-negligible non-diagonal terms due to the spatial dispersion effect \cite{grapheneModeling3,carbonNanotubes}. %%
Additionally, in the presence of a magnetic bias field, a similar dyadic conductivity is needed to model the graphene sheet. %%
Therefore, the conductivity is generally modeled as the following tensor: %%
\begin{equation}
\label{tensorConductivityModel}
\underline{\vec{\sigma}} ( \omega , \mu_c (\mathbf{E}_0) , \Gamma , T , \mathbf{B}_0 )  = \left[ \begin{array}{cc} \sigma_{xx} & \sigma_{xy} \\ \sigma_{yx} & \sigma_{yy} \end{array} \right]
\end{equation}
where $\omega$ is the radial frequency, $\mu_c$ is the chemical potential hinging upon the applied electrostatic bias field $\mathbf{E}_0 = \unitvec{z} E_0$ or doping, $\Gamma$ is a phenomenological electron scattering rate, $T$ is the temperature and $\mathbf{B}_0 = \unitvec{z} B_0$ is the applied magnetostatic bias field. %%
In this study, we consider graphene periodic surfaces under electric bias field only, thus the anisotropic effect emanating from magnetic biasing is set to zero.
The four elements of the conductivity tensor can be written in the following general form \cite{grapheneModeling3}: %%
\begin{align}
\label{tensorConductivityElements}
\displaystyle \sigma_{xx} & = \alpha \frac{\partial^2}{\partial x^2} + \beta \frac{\partial^2}{\partial y^2} + \sigma \nonumber \\
\displaystyle \sigma_{xy} & = 2 \beta \frac{\partial^2}{\partial x \partial y} \nonumber \\
\displaystyle \sigma_{yx} & = 2 \beta \frac{\partial^2}{\partial x \partial y} \nonumber \\
\displaystyle \sigma_{yy} & = \beta \frac{\partial^2}{\partial x^2} + \alpha \frac{\partial^2}{\partial y^2} + \sigma.
\end{align}
The operator terms in \eqref{tensorConductivityElements} represent the spatial dispersion effect and are negligible either in low frequency regime or high phase velocities. %%

The coefficients in \eqref{tensorConductivityElements}, namely $\alpha$, $\beta$, and $\sigma$, can be obtained experimentally or analytically based on the existing models. %%
In fact, the theoretical modeling of graphene conductivity has been the subject of many recent research efforts, including successful comparisons with experimental results \cite{grapheneTheoreticalModeling1,grapheneTheoreticalModeling2,grapheneTheoreticalModeling3,grapheneTheoreticalModeling4,grapheneConductivityModel}. %%
However, like any other material properties, these coefficients are strongly influenced by the fabrication process and environmental effects. %%
For instance, the defects introduced by the substrate in the graphene sheet often incur deviations from the theoretical predictions \cite{grapheneSubstrateEffect}.
In the presented study, the numerical routine is developed for general parameters and later in the numerical examples values obtained based on theoretical models are considered. %%

\begin{figure}[!t]
\centering
\subfloat[]{\includegraphics[width=3.375in]{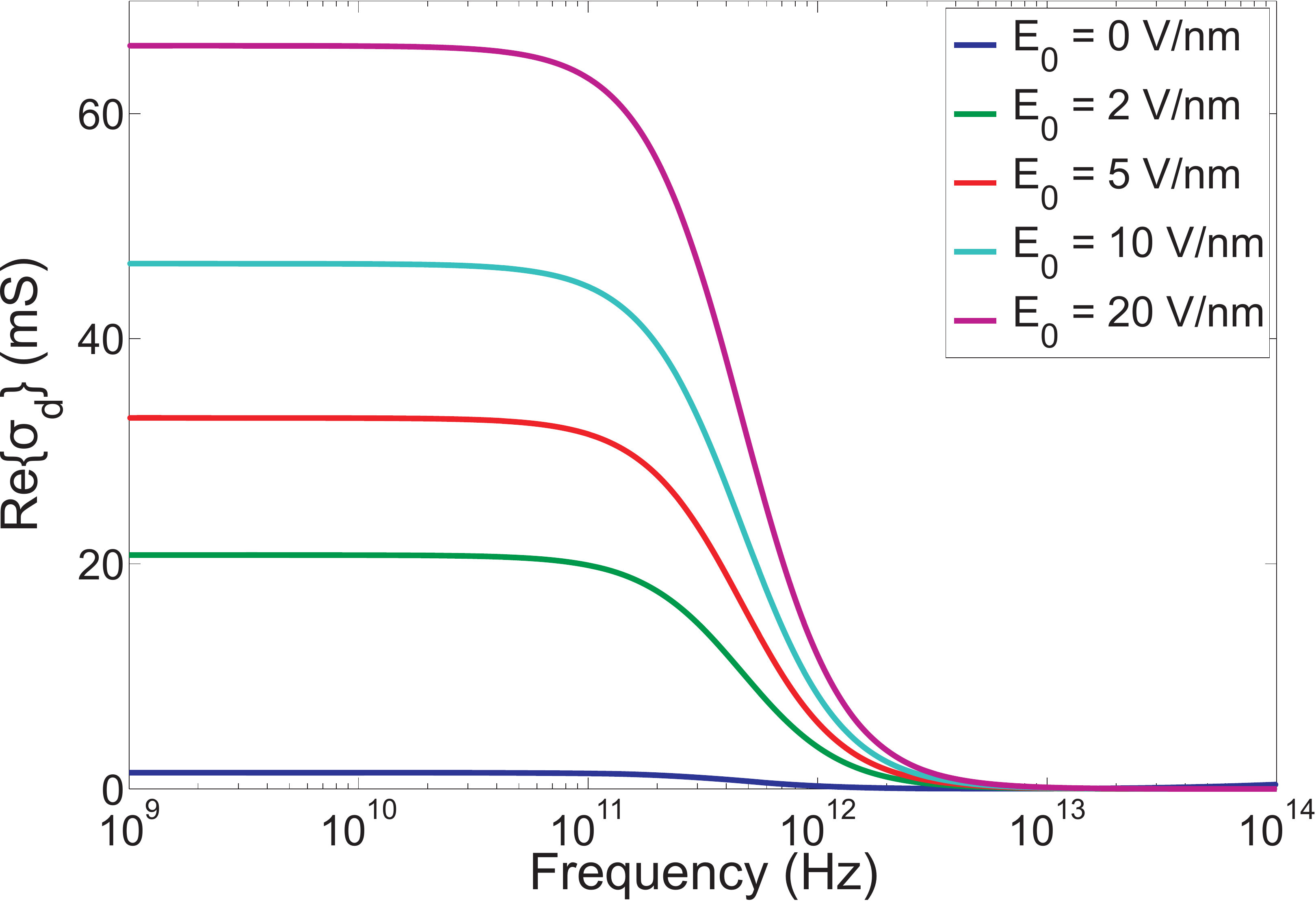} \label{conductivityReal}}
\\
\subfloat[]{\includegraphics[width=3.375in]{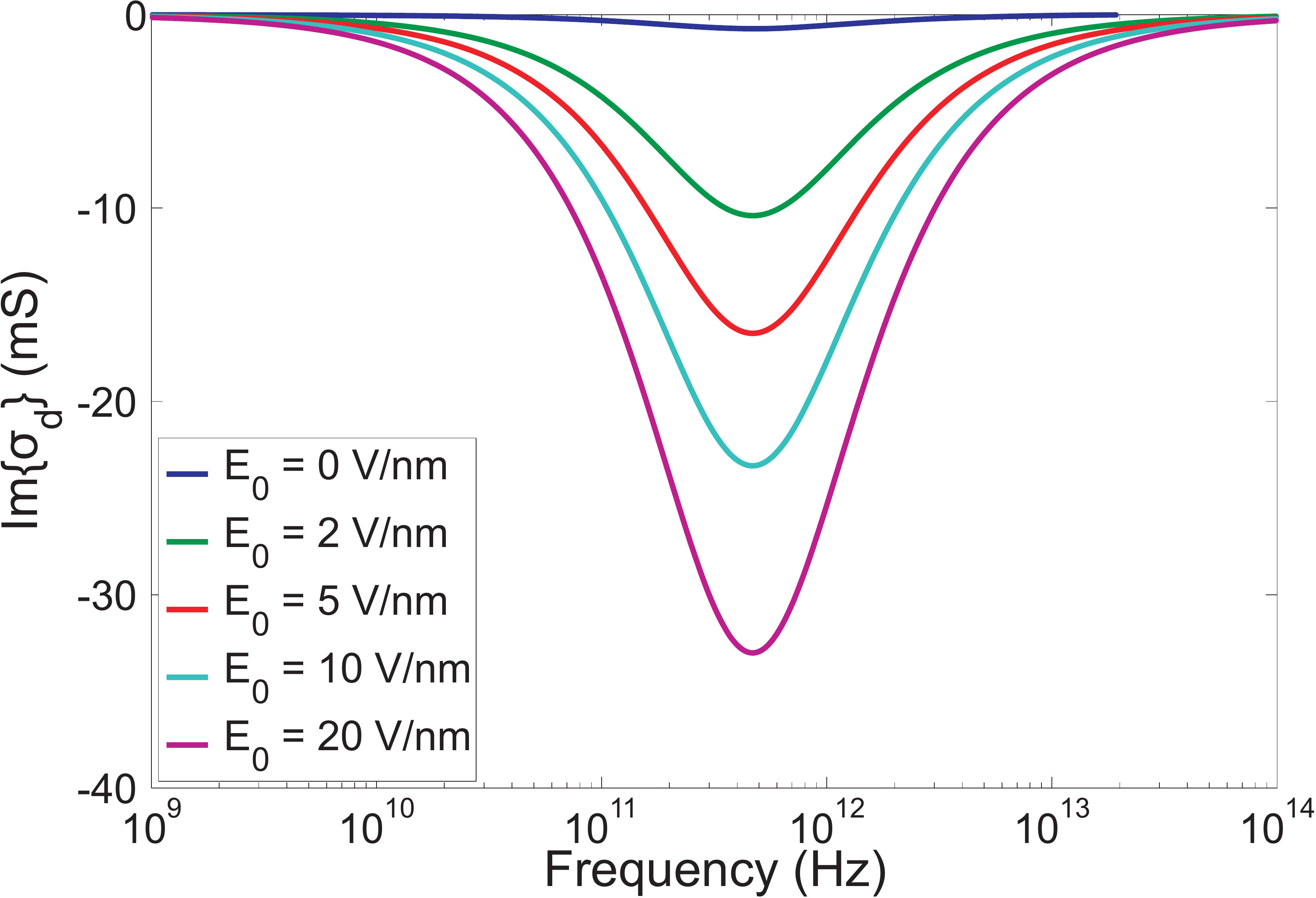} \label{conductivityImag}}
\caption{(a) Real part and (b) imaginary part of the conductivity ($\sigma_d$) in terms of frequency for various bias electric field in microwave regime}
\label{conductivity}
\end{figure}

The conductivity value $\sigma$ can be theoretically calculated using Kubo's formalism \cite{grapheneConductivityModel}, which yields the following equation:
\begin{align}
\label{conductivityGrapheneReal}
\displaystyle \sigma & ( \omega , \, \mu_c (\mathbf{E}_0) , \Gamma , T , \mathbf{B}_0 ) = \frac{e^2 v_F^2 |eB_0| (\omega-j2\Gamma) \hbar}{-j \pi} \sum\limits_{n=0}^{\infty} \nonumber \\
& \left\{ \frac{f_d(M_n) - f_d(M_{n+1})+f_d(-M_{n+1})-f_d(-M_n)}{(M_{n+1}-M_n)^2 - (\omega - j2\Gamma)^2\hbar^2} \right. \nonumber \\
& \times \frac{1-\Delta^2 / (M_n M_{n+1})}{M_{n+1} - M_n} \nonumber \\
& + \frac{f_d(-M_n) - f_d(M_{n+1})+f_d(-M_{n+1})-f_d(M_n)}{(M_{n+1}+M_n)^2 - (\omega - j2\Gamma)^2\hbar^2} \nonumber \\
& \left. \times \frac{1+\Delta^2/(M_n M_{n+1})}{M_{n+1} + M_n} \right\}
\end{align}
where $M_n = \sqrt{ \Delta^2 + 2nv_F^2|eB_0|\hbar}$, and $f_d(\varepsilon)$ is the Fermi-Dirac distribution given by
\begin{equation}
\label{fermiDiracDistribution}
f_d(\varepsilon) = \frac{1}{1+e^{(\varepsilon-\mu_c)/(k_BT)} }.
\end{equation}
The above equation returns the conductivity for a general case, with both electric and magnetic bias fields. %%
However, for the case of no magnetic bias field, the above equation should be calculated when $|eB_0|$ tends to zero. %%
In this case, the following equation should be used \cite{grapheneModeling2}:
\begin{align}
\label{conductivityGrapheneNoMagneticBias}
\displaystyle \sigma & ( \omega , \, \mu_c (\mathbf{E}_0) , \Gamma , T ) = \frac{j e^2 (\omega-j2\Gamma)}{\pi \hbar^2} \nonumber \\
& \displaystyle \left\{ \frac{1}{(\omega-j2\Gamma)^2} \int_0^{\infty} \varepsilon \left( \frac{\partial f_d (\varepsilon)}{\partial \varepsilon} - \frac{\partial f_d (-\varepsilon)}{\partial \varepsilon} \right) d\varepsilon \right. \nonumber \\
& \displaystyle \left. - \int_0^{\infty} \frac{ f_d (-\varepsilon) - f_d (\varepsilon)}{(\omega-j2\Gamma)^2-4(\varepsilon/\hbar^2)} d\varepsilon \right\}
\end{align}
Room temperature ($T=300\,\mbox{K}$) is considered throughout the paper. %%
Correspondingly, the excitonic energy gap $\Delta$ is set to zero, the Fermi velocity in graphene is $v_F=10^6\,\mbox{m/s}$ and $\Gamma$ is set to 12.2 meV \cite{masslessDiracFermions4,grapheneSubstrateEffect,GammaValue}. %%
Note that for the case of no magnetic bias field, the above equation should be calculated when $|eB_0|$ tends to zero. %%
Fig.\,\ref{conductivity} shows the variation of graphene conductivity with frequency for different bias electrostatic fields in the microwave and infrared regime.
In the microwave regime, though the conductivity does not change significantly with frequency, it is strongly dependent on the bias electric field. %%
The curves evidence stronger sensitivity of the conductivity to the bias field at low THz frequencies compared to the higher ones. %%
It can be observed that changing the bias electrostatic field values directly affects the range of frequency over which graphene behaves as a conductor.

The coefficients $\alpha$ and $\beta$ are obtained using a perturbation theory approach presented in \cite{grapheneModeling3}. %%
Although this was initially made for unbiased graphene, following the same approach it can be shown that the final results also hold for electrically
biased graphene. %%
The equations providing the values of $\alpha$ and $\beta$ then reads as
\begin{equation}
\label{spatialDispersionCoefficients}
\alpha = \frac{3}{4}\frac{v_F^2}{(\omega-j2\Gamma)^2} \qquad \mbox{and} \qquad \beta=\frac{\alpha}{3}.
\end{equation}

\section{PMoM for Graphene Biperiodic Surfaces}

Modeling EM properties of homogeneous graphene layers has been targeted in some previous publications \cite{grapheneModeling1,grapheneModeling2,grapheneModeling3,grapheneModeling4}, where the solution of the Maxwell equations in a one-dimensional space is used to simulate multilayer structures. %%
The analysis of periodically patterned graphene demands more advanced analysis techniques, to account for the the periodic boundary condition on the electromagnetic fields as well as the boundary condition on the induced currents, i.e. normal component at the graphene boundaries should vanish. %%
The properties of an EM screen can be obtained by formulating the scattered field as a result of a plane wave excitation. %%
The whole metasurface characteristics, such as resonance frequencies, high absorption frequency points, band diagram and polarization sensitivity, are then extracted from the scattered field spectrum. %%
Here a periodic method of moments (PMoM) technique is implemented, which efficiently solves Maxwell's equations for multilayer planar metasurfaces \cite{myThesis}.
This full-vector semi-analytical numerical technique not only allows accurately modeling periodic graphene surfaces but also includes the effect of the substrate actual topology. %%

The concept of PMoM for periodic surfaces of various types has been introduced in many publications \cite{bookFSS1,bookFSS2,myThesis}. %%
The method begins with the formulation of the boundary condition which must hold on the graphene surface, i.e. %%
\setlength{\arraycolsep}{0.14em}
\begin{equation}
\label{boundarycondition} \left [ \begin{array}{c} E_x^i \\
E_y^i \end{array} \right]+\left [ \begin{array}{c} E_x^s \\
E_y^s \end{array} \right]=-Z_s \left [ \begin{array}{c} J_x \\
J_y \end{array} \right],
\end{equation}
where $\mathbf{E}_t^s=[E_x^s,E_y^s]^T$ and $\mathbf{E}_t^i=[E_x^i,E_y^i]^T$ are tangential components of the scattered and incident electric field on graphene, respectively. %%
In addition, $Z_s$ stands for the surface impedance and $\mathbf{J}=[J_x,J_y]^T$ is the induced current on the graphene surface. %%
The scattered field can be written in terms of a Green's function and the induced currents. %%
In case of a homogeneous substrate, this yields the following equation: %%
\begin{align}
\label{homocase}
 -\left[ \begin{array}{c}
 E_x^i \\
 E_y^i \end{array} \right] = & \sum_{m=-\infty}^{\infty} \sum_{n=-\infty}^{\infty} \left( \left[ \begin{array}{cc}
 \tilde{G}_{xx_{mn}} &  \tilde{G}_{xy_{mn}} \\
 \tilde{G}_{yx_{mn}} &  \tilde{G}_{yy_{mn}} \end{array} \right] + \left[ \begin{array}{cc}
 Z_s & 0 \\
 0 &  Z_s \end{array} \right] \right) \nonumber \\
 & \cdot \left[ \begin{array}{c}
 \tilde{J}_{x_{mn}} \\
 \tilde{J}_{y_{mn}} \end{array} \right] e^{-(jk_{xm} x+jk_{yn} y)}
\end{align}
where $\tilde{G}_{xx}$, $\tilde{G}_{xy}$, $\tilde{G}_{yx}$ and $\tilde{G}_{yy}$ are the components of the dyadic Green's function in the spectral domain. %%
$k_{xm}$ and $k_{yn}$ are given by %%
\begin{equation}
\label{alphambetan}
\displaystyle k_{xm}=\frac{2\pi m}{L_x}+k_x \qquad \mbox{and} \qquad
\displaystyle k_{yn}=\frac{2\pi n}{L_y}+k_y
\end{equation}
where $\vec{\mathbf{k}}_\mathbf{{inc}}=k_x\hat{x}+k_y\hat{y}$ is the wave vector of the incident plane wave and the pair $(L_x,L_y)$ determines the lattice constants of the periodic structure in both $x$ and $y$ directions, respectively.

If the summations in \eqref{homocase} are cast in a matrix form, the following equation is obtained: %%
\setlength{\arraycolsep}{0.14em}
\begin{equation}
\label{matrixform1} -\left[ \begin{array}{c}
E_x^i \\
E_y^i \end{array} \right]_{(x,y)} = \mathbf{A}_{(x,y)} \left(
\tilde{\mathbf{G}} + \mathbf{Z}_s \right) \left[ \begin{array}{c}
\tilde{\mathbf{J}}_x \\
\tilde{\mathbf{J}}_y
\end{array} \right]
\end{equation}
with
\begin{equation}
\mathbf{A}_{(x,y)} = \left [ \begin{array}{cc} \left[e^{ \; (jk_{xm} x+jk_{yn} y)} \right]^T & \left[0\right]^T \\
\left[0\right]^T & \left[e^{ \; (jk_{xm} x+jk_{yn} y)}\right]^T
\end{array} \right ].
\end{equation}
$[\exp(jk_{xm} x+jk_{yn} y)]^T$ is a row matrix containing the exponential terms and $[0]^T $ is a zero matrix with the same size as $[\exp(jk_{xm} x+jk_{yn} y)]^T$. %%
In fact, the superscript $T$ representing the transpose sign is used to distinguish between row and column vectors. %%
$\tilde{\mathbf{J}}_x$ and $\tilde{\mathbf{J}}_y$ are column vectors obtained from the Fourier coefficients $\tilde{J}_{x_{mn}}$ and $\tilde{J}_{y_{mn}}$ in $(k_{xm},k_{yn})$ basis. %%
Finally, $\tilde{\mathbf{G}}$ is the Green's function matrix.

For solving (\ref{matrixform1}) using the concept of the MoM, electric currents excited on the patches should be expanded in terms of some basis functions %%
\begin{equation}
\label{matrixform2}
\left [ \begin{array}{c} J_x \\ J_y
\end{array} \right ] = \left [
\begin{array}{c} \mathbf{J}_x^T \\ \mathbf{J}_y^T
\end{array} \right ] e ^ {j \mathbf{k}_\mathrm{inc} \cdot
\mathbf{r}} \cdot \mathbf{C}
\end{equation}
where $\mathbf{J}_x^T$ and $\mathbf{J}_y^T$ are row vectors containing the basis functions used for expanding $J_x$ and $J_y$, respectively. %%
The unknown coefficients of these functions are arranged in the column vector $\mathbf{C}$. %%
Using Galerkin's method and after some algebraic operations the following system of equations is obtained: %%
\begin{align}
\label{PMoM}
\displaystyle - \left [ \int \mathbf{J}^* e ^ {j \mathbf{k}_\mathrm{inc} \cdot
\mathbf{r}} \cdot \mathbf{E}^i \, \mathrm{d}s \right ] = &
\left [ [ \tilde{\mathbf{J}}_x ]^\dag \,\, [ \tilde{\mathbf{J}}_y ]^\dag \right ]
( \tilde{\mathbf{G}} + \mathbf{Z}_s ) \nonumber \\
& \cdot \left[ \begin{array}{c} {} [\tilde{\mathbf{J}}_x] \\ {} [\tilde{\mathbf{J}}_y]
\end{array} \right] \mathbf{C}
\end{align}
where $\mathbf{J}=\mathbf{J}_x \hat{x}+\mathbf{J}_y \hat{y}$ and $\mathbf{E}^i=E^i_x \hat{x} +E^i_y \hat{y}$ is the incident electric field vector. %%
$[ \tilde{\mathbf{J}}_x ]$ and $[ \tilde{\mathbf{J}}_y]$ are matrices whose \emph{k}'th columns are Fourier coefficients of \emph{k}'th corresponding basis functions. %%
The signs $^*$ and $^\dag$ stand for the complex and Hermitian conjugate respectively. %%
$\tilde{\mathbf{J}}_x$ and $\tilde{\mathbf{J}}_y$ in (\ref{matrixform1}) are related to $[ \tilde{\mathbf{J}}_x ]$ and $[\tilde{\mathbf{J}}_y ]$ through %%
\begin{equation}
\label{currentExpansion}
\left[ \begin{array}{c} \tilde{\mathbf{J}}_x \\ \tilde{\mathbf{J}}_y
\end{array} \right] = \left [ \begin{array}{c} [ \tilde{\mathbf{J}}_x ] \\ {} [
\tilde{\mathbf{J}}_y ] \end{array} \right ] \cdot \mathbf{C}.
\end{equation}
Using the obtained coefficients $\mathbf{C}$, all the desired quantities such as reflection and transmission coefficients can easily be calculated.
The term $\exp(j \mathbf{k}_\mathrm{inc} \cdot \mathbf{r})$ is considered as a phase factor in all basis functions. Therefore, the Fourier coefficients are calculated in $(\frac{2m\pi}{L_x},\frac{2n\pi}{L_y})$ basis.

The studies on modeling planar geometries containing periodic patches are mainly limited to considering various substrate properties. %%
Different schemes are developed for homogeneous, lossy, multilayer \cite{bookFSS1,bookFSS2}, periodic \cite{myThesis} and anisotropic \cite{anisoFSSanalysis1,anisoPeriodicFSSAnalysis} substrates. %%
Thin metallic patches are accurately modeled with an equivalent scalar surface impedance. %%
Therefore, a scalar value is always assumed for the matrix $\mathbf{Z}_s$.
However, for graphene patches, a more general form should be developed for $\mathbf{Z}_s$, which is actually the goal in this section.

Let us first take the conductivity equation $\mathbf{J}_t = \underline{\vec{\sigma}} \mathbf{E}_t$ into account. %%
By considering the operator form of the graphene conductivity, the general equation for graphene reads as
\begin{equation}
\label{conductivityEquation}
\left[ \begin{array}{c} J_x \\ J_y \end{array} \right]=
\left[ \begin{array}{cc}
\sigma + \alpha \frac{\partial^2}{\partial x^2} + \beta \frac{\partial^2}{\partial y^2} &
2 \beta \frac{\partial^2}{\partial x \partial y} \\
2 \beta \frac{\partial^2}{\partial x \partial y} &
\sigma + \beta \frac{\partial^2}{\partial x^2} + \alpha \frac{\partial^2}{\partial y^2}
\end{array} \right]
\left[ \begin{array}{c} E_x \\ E_y \end{array} \right].
\end{equation}
For the analysis of a frequency selective surface using PMoM, this equation should be transformed in the spectral domain. %%
For this purpose, the equivalence equations $\partial / \partial x \equiv -j \vec{k_x}$ and $\partial / \partial y \equiv -j \vec{k_y}$ should be used, where $\vec{k_x}$ and $\vec{k_y}$ are diagonal matrices with diagonal elements equal to $k_{xm}$ and $k_{yn}$. %%
Therefore, the conductivity equation for graphene in the spectral domain is as the following:
\begin{equation}
\label{conductivityEquationSpectral}
\left[ \begin{array}{c} \tilde{\mathbf{J}}_x \\ \tilde{\mathbf{J}}_y \end{array} \right]=
\left[ \begin{array}{cc} \displaystyle
\vec{\sigma} - \alpha \vec{k_x}^2 - \beta \vec{k_y}^2 &
\displaystyle - 2 \beta \vec{k_x} \vec{k_y}  \\
\displaystyle - 2 \beta \vec{k_x} \vec{k_y}  &
\displaystyle \vec{\sigma} - \beta \vec{k_x}^2 - \alpha \vec{k_y}^2
\end{array} \right]
\left[ \begin{array}{c} \tilde{\mathbf{E}}_x \\ \tilde{\mathbf{E}}_y \end{array} \right],
\end{equation}
By simply comparing \eqref{conductivityEquationSpectral} with \eqref{matrixform1}, the matrix $\mathbf{Z}_s$ can be deduced as %%
\begin{equation}
\label{impedanceMatrixGraphene}
\mathbf{Z}_s =
\left[ \begin{array}{cc} \displaystyle
\vec{\sigma} - \alpha \vec{k_x}^2 - \beta \vec{k_y}^2 &
\displaystyle  - 2 \beta \vec{k_x} \vec{k_y}  \\
\displaystyle  - 2 \beta \vec{k_x} \vec{k_y}  &
\displaystyle \vec{\sigma} - \beta \vec{k_x}^2 - \alpha \vec{k_y}^2
\end{array} \right] ^ {-1}.
\end{equation}
Note that the inverted matrix consists of four diagonal matrices. %%
Thus, its inversion can be accomplished analytically, without the need to follow complicated computational procedures for obtaining $\mathbf{Z}_s$. %%
Once this impedance matrix is evaluated, it is plugged in \eqref{matrixform1} and the common process for solving the metasurface problem is carried out. %%
The presented formulation is a general procedure allowing to solve unbiased and biased graphene metasurfaces under plane wave incidence. %%
However, the examples considered next do not include magnetic bias thus there is no non-diagonal terms due to magnetic field. %%
Concerning the contribution of spatial dispersion to the non-diagonal conductivity terms, it is negligible in our examples (see below) and thus the conductivity writes in essence: %%
\begin{equation}
\label{surfaceImpedance}
\mathbf{Z}_s = \left[ \begin{array}{cc} \sigma & 0 \\ 0 & \sigma \end{array} \right].
\end{equation}

\section{Application Examples}

The remainder of the paper presents different examples of tunable biperiodic graphene metasurfaces, modeled and designed using the PMoM for graphene described in the previous sections. %%
We begin with a simple fundamental case to illustrate the physical basis of the considered configurations and their potential. %%
Then, more realistic structures from both nanofabrication and biasing control point of views are considered. %%
Note that the effect of edges on graphene conductivity is also neglected, since the smallest dimensions considered are in the order of $1\mu \mbox{m}$, which is much larger than the electron scattering length within the graphene layer.

\subsection{Freestanding metasurface}

\begin{figure}[!t]
\centering
\includegraphics[width=3.375in]{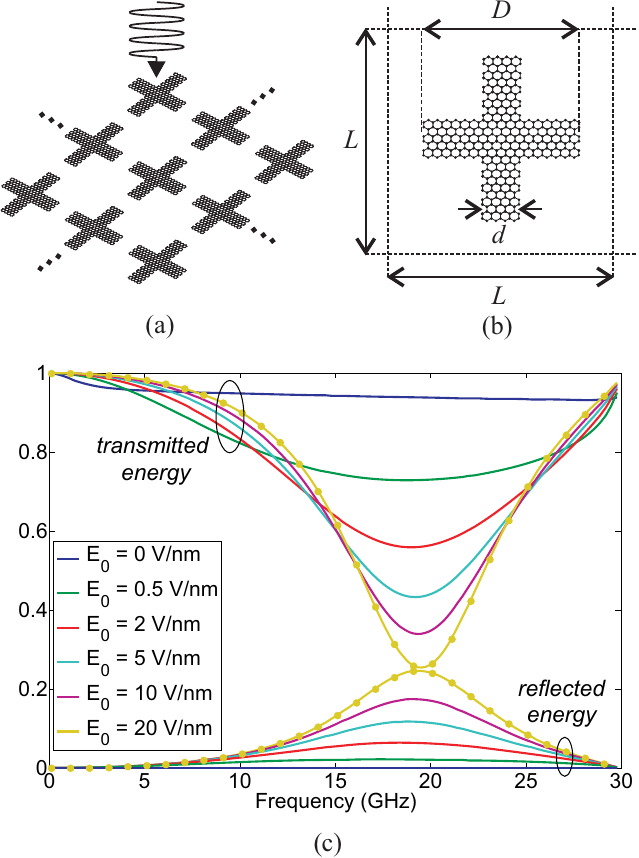}
\caption{(a) Schematic illustration of the graphene biperiodic structure consisting of graphene cross shaped patches arranged in a two dimensional lattice and under a normally incident plane wave. (b) Unit cell of the lattice. (c) Reflected and transmitted energy versus frequency. The filled circles represent the results obtained when accounting for the spatial dispersion terms in the case $E_0=20\,\mbox{V/nm}$.}
\label{crossShapePatch}
\end{figure}

\begin{figure}[!t]
\centering
\includegraphics[width=3.375in]{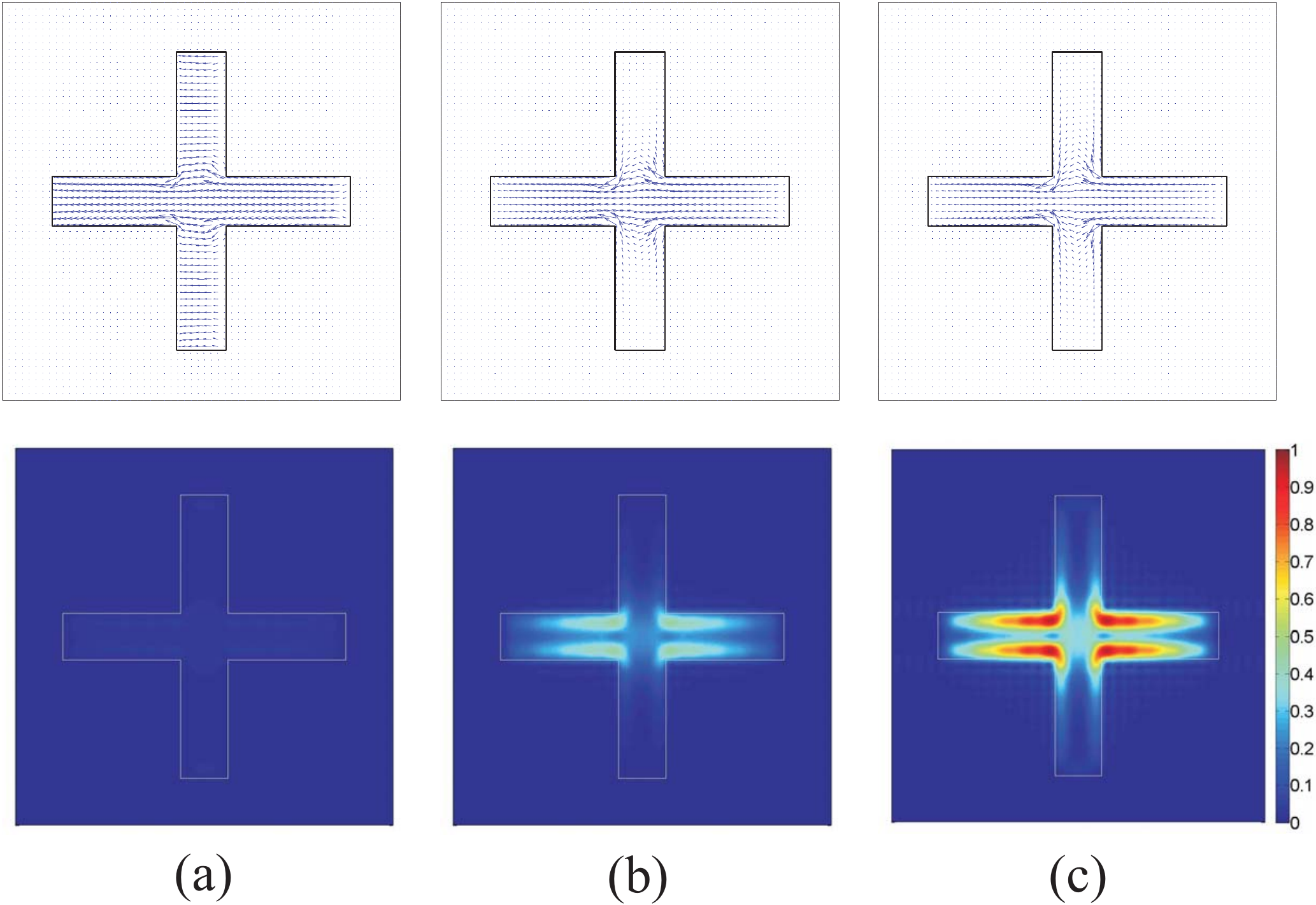}
\caption{Normalized induced current on the patch at $\omega=19.5\,\mbox{GHz}$ for different applied electrostatic fields are plotted for a free standing FSS consisting of 2D lattice of cross shaped graphene patches exposed to a normal incident plane wave: (a) $E_0=0\,\mbox{V/nm}$, (b) $E_0=2\,\mbox{V/nm}$ and (c) $E_0=20\,\mbox{V/nm}$ }
\label{currentProfiles}
\end{figure}

The simplified topology, considered first, consists of a single biperiodic surface, whose unit cell is a \emph{cross shaped} graphene patch (Fig.\,\ref{crossShapePatch}a and \ref{crossShapePatch}b). %%
In this example, results are computed assuming that the patterned graphene layer is free-standing in air or vacuum and the effect of gating on its performance is studied. %%
The dimensions of the patch are $L=10 \, \mbox{mm}$, $d=1.25\, \mbox{mm}$, $D=7.5 \, \mbox{mm}$ and a normal incidence of the plane wave on the graphene layer is assumed. %%
The frequency range for the analysis is $0\,\mbox{GHz} < f < 30\,\mbox{GHz}$. %%
Fig.\,\ref{crossShapePatch}c shows the reflected and transmitted energy plotted in terms of the excitation frequency for various bias electric field values.
The case $E_0=20\,\mbox{V/nm}$ is also analyzed including the spatial dispersion terms, which confirms their negligible effect.
The results reveal an extremely promising feature, which is possibility to electrically control the \emph{emergence} and strength of resonances. %%
Note that this capability is different from that presented by Ju et al. \cite{grapheneStripes2}, since there only the \emph{position} of the resonance is controlled.
Indeed, applying the bias electric field results in a strong increase in the surface conductivity, thereby exciting the resonances between adjacent graphene patches.
These resonances introduce maximum and minimum points in the reflection and transmission spectra. %%
This behavior can be verified by visualizing the current profile induced on the graphene patches at $f=19.5\,\mbox{GHz}$, as done in Fig.\,\ref{currentProfiles} for three different electrostatic gatings, namely $E_0=0\,\mbox{V/nm}$, $E_0=2\,\mbox{V/nm}$ and $E_0=20\,\mbox{V/nm}$.
It is observed that the induced currents gradually take on a resonating form when the bias electric field increases, meaning that the current density sharply increases at the center of the patch and vanishes at the edges. %%
Though this illustration was mainly devised and presented for a first physical insight on controllable graphene metasurface, it is noticeable that the demonstrated switchable frequency selective property would find real applications. %%
For instance, radar absorbers can be realized, where the absorption peaks are switched on and off through an external voltage. %%

As mentioned earlier, this initial illustration is presented merely to introduce the advantage of graphene patterning and does not constitute a realistic electrically biased graphene metasurface. %%
From an experimental point of view, the graphene layer must reside on a substrate and for simple biasing it is impractical to have DC-disconnected graphene unit cells. %%
Graphene layers are sensitive to normal bias electrostatic fields in the order of $10^9\,\mbox{V/m}$. %%
Even relatively thin dielectric layer spacer can result in very large bias voltages to achieve the required electrostatic bias \cite{grapheneMetamaterials4}. %%
Gate voltages in the order of maximum 100\,V-200\,V thus require a very thin dielectric between graphene layer and gate, namely, in the order of some tens of nanometers. %%
Such thin dielectrics are implementable and can be employed in most applications of graphene to realize nanoelectronic devices. %%
However, they become problematic when considering graphene layers affecting the propagation of an EM wave, since a metal or silicon conductive gate layer in close proximity to the graphene layer would generally mask - or at least strongly affects - the desired tunable property. %%
Therefore, the remainder of the paper considers metasurfaces including actual substrates, and where the different unit cells of a graphene layer are DC-connected so as to only require applying DC biasing at single location on the surface. %%
The maximum DC bias voltage considered is 200V. %%

For this purpose, we propose to use periodic metasurfaces made of different layers of graphene \cite{grapheneModulators}. %%
Such a stack can be realized by first transferring a uniform graphene layer on the substrate and patterning it via e-beam lithography to enable geometrical features down to a few tens of nanometers.  %%
Then, the thin dielectric is deposited and finally the transferring and patterning steps are repeated for the second graphene layer. %%
In fact, the process can also be simplified by etching all graphene and dielectric layers in a single step. %%
This topology has two main advantages: 1) It provides more freedom for tailoring the surface EM response and 2) it allows an efficient dynamic bias of the structure by applying a voltage between the graphene layers themselves as symbolized in Fig.\,\ref{dcConnectedPatch}b. %%
Hence, the aforementioned masking of the graphene metasurface by a neighboring metal or silicon gate is basically avoided. %%
Obviously, a DC contact between the lower layer and the voltage source must be implemented at one location in the surface, but this is also very simply realized through an access cavity in the thin dielectric or implementing vias.  %%
Furthermore, each patch of the periodic metasurface must be connected to the neighboring ones for transferring the DC voltage throughout the whole surface. %%
Therefore, owing to the conductive properties of graphene, a periodic structure with DC connected patches is preferred. %%
In this case, no additional circuit or layer to support for the DC gating is required. %%

\subsection{DC-connected double layer metasurface}

\begin{figure}[t]
\centering
\includegraphics[width=3.375in]{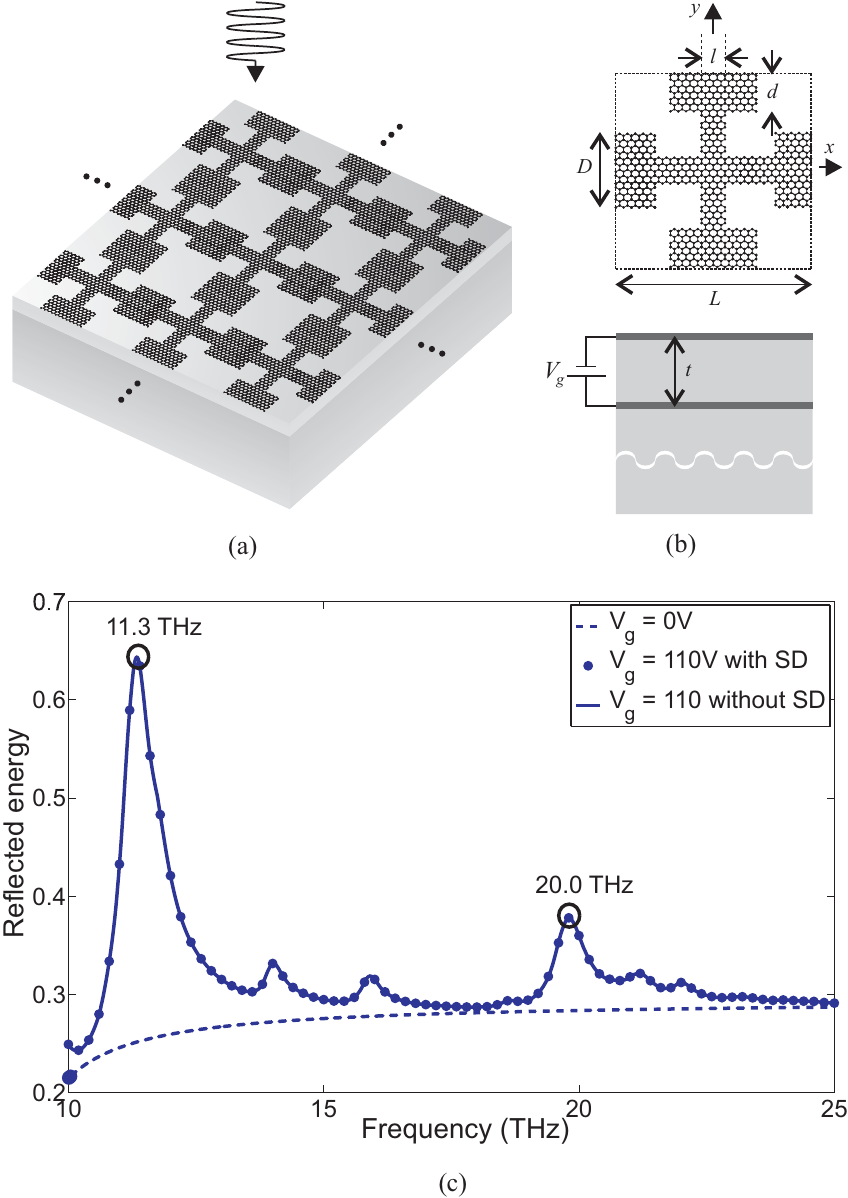}
\caption{(a) Schematic illustration of the graphene biperiodic structure consisting of DC connected patches on both sides of a GaAs layer residing on an infinitely thick GaAs substrate and assumed to be subjected to a normally incident plane wave. (b) The unit cell of the lattice is depicted with the dimensions; the top figure is the top view of the unit cell and the bottom figure is its side-view. (c) The reflected energy versus frequency is plotted for unbiased case as well as biased electrostatic field with $E_0=2.2\,\mbox{V/nm}$.}
\label{dcConnectedPatch}
\end{figure}

\begin{figure}[t]
\centering
\includegraphics[width=3.375in]{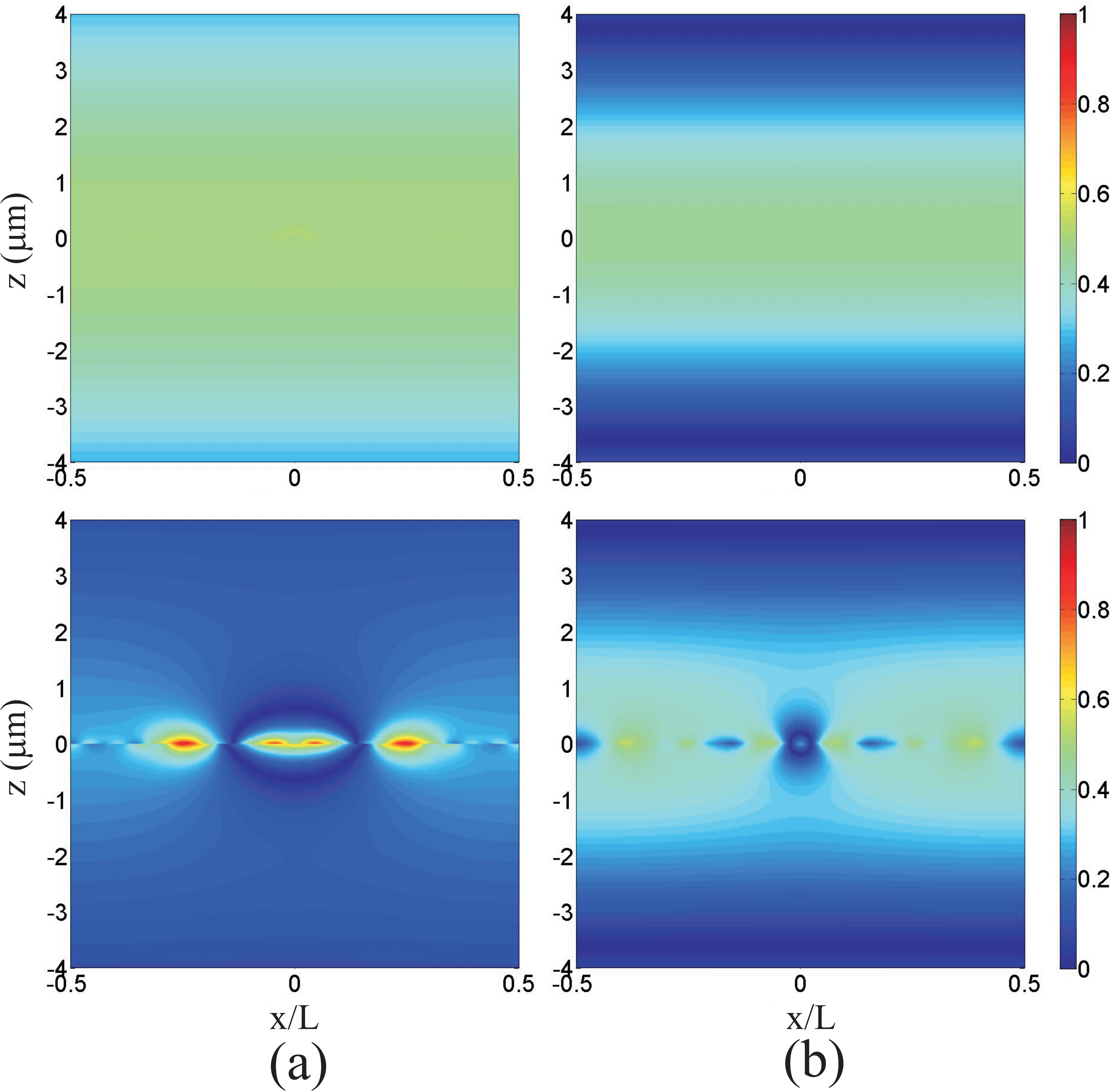}
\caption{Magnitude of the total electric field is shown for two resonant frequencies, namely (a) 11.3\,THz and (b) 20.0\,THz. The upper figures correspond to unbiased graphene patches and the lower ones are obtained for an electrically biased graphene surface with $E=2.2\,\mbox{V/nm}$.}
\label{fieldProfile}
\end{figure}

Based on the described technological approach, a graphene metasurface was designed to operate as a frequency selective surface in the far infrared range, to demonstrate the potential of the proposed concept at higher frequencies. %%
The considered surface and unit cell are depicted in Fig.\,\ref{dcConnectedPatch}a and \ref{dcConnectedPatch}b. %%
The same pattern is etched in both graphene layers, with $L=5\,\mu\mbox{m}$, $l=0.5\,\mu\mbox{m}$, $d=1.25\,\mu\mbox{m}$, and $D=1.5\,\mu\mbox{m}$. %%
The two patch layers sandwich a $L=50\,\mbox{nm}$ thick GaAs layer over an infinitely thick GaAs substrate. %%
The relative permittivity of GaAs at DC is equal to $\epsilon_r^{\mathrm{DC}}=10.8$ and at far infrared frequencies measured data for the dispersive dielectric constant is considered \cite{GaAsOpticalMeasurement}.
Note that though GaAs might not be the best dielectric for graphene, in fact the performance computed here are weakly dependent on the dielectric permittivity. %%
In any case, the results could be easily updated for other materiel choices, such as Parylene having similar permittivities.
A 110V bias voltage, corresponding to the normal electrostatic field $E_0=2.2\,\mbox{V/nm}$, is applied between the two graphene layers. %%
In Fig.\,\ref{dcConnectedPatch}c, the reflected energy under the excitation of a $x$-polarized plane wave is shown, when the bias voltage is switched on and off. %%
The sharp resonances seen in the reflection spectrum occurs due to the increase in the graphene conductivity. %%
The devised graphene biperiodic layer constitute a controllable ultra-thin multiple band filter at infrared frequencies. %%
The device simulation is carried out with and without consideration of spatial dispersion, which confirms the negligible effect of spatial dispersion in this example as well.
In Fig.\,\ref{fieldProfile}, the magnitude of the electric field in the $xz$ plane is shown at the two different resonance frequencies. %%
Two different cases of biased and unbiased graphene patches are considered. %%
The results evidence the considerable variation of the transmitted field with the electrostatic gating, which is more pronounced in the first resonance. %%

\subsection{Graphene microwave absorber}

\begin{figure}[t]
\centering
\includegraphics[width=3.375in]{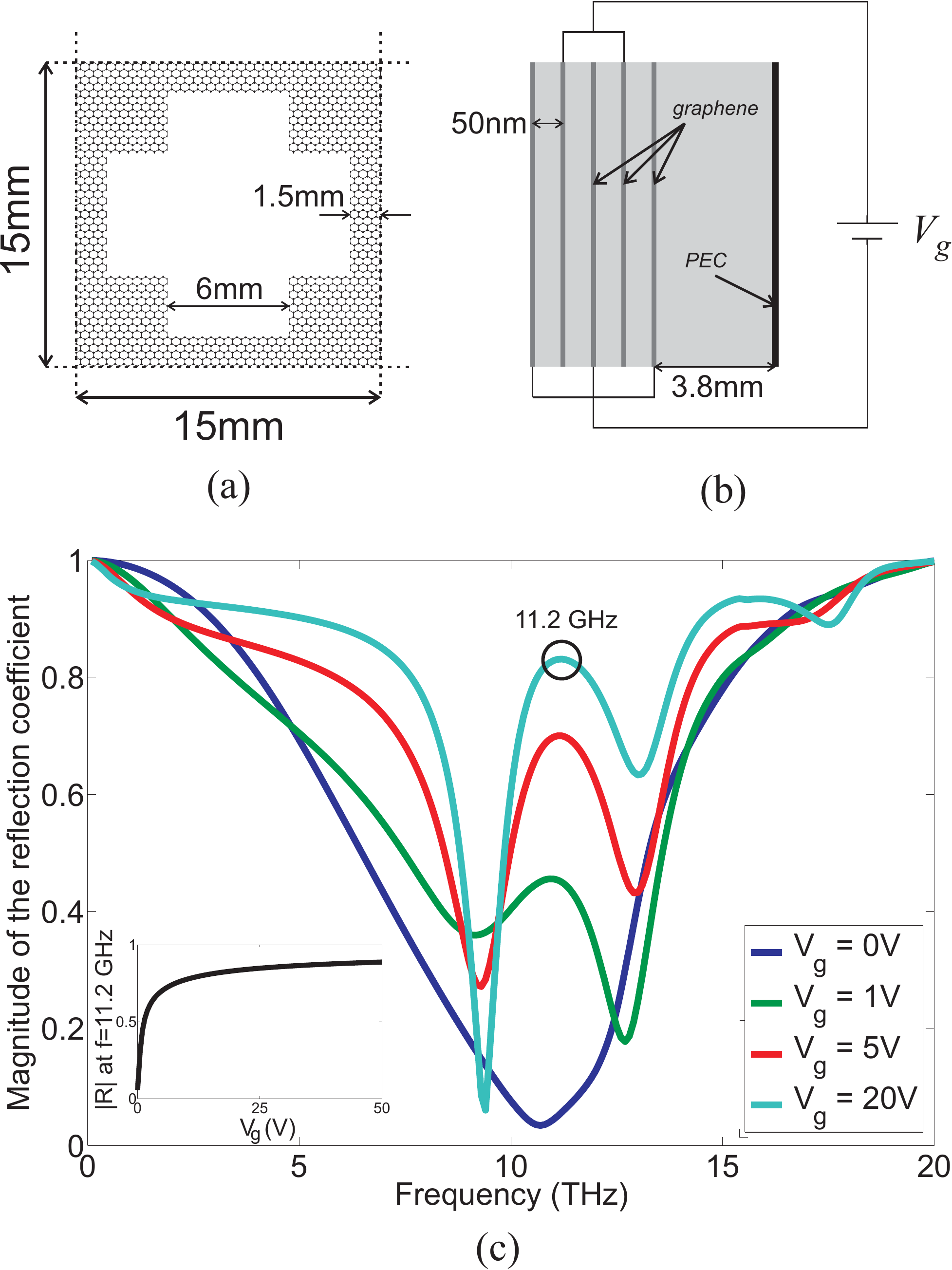}
\caption{Schematic illustration of the absorber structure consisting of five graphene layers printed on a glass substrate: (a) The graphene patch layer, (b) the side-view of the absorber and the biasing circuit. (c) Magnitude of the reflection coefficient versus frequency for different applied bias voltages ($V_g$) are plotted. In the subfigure, magnitude of the reflection coefficient at $f_0=11.2\,\mbox{GHz}$ is plotted in terms of the applied bias voltages ($V_g$) for the considered absorber.}
\label{microwaveAbsorber}
\end{figure}

A general conclusion from the presented analyses is that using graphene technology ultrathin surfaces can be devised, which influence dramatically the incident wave. %%
Moreover, in these extremely thin geometries, one has the possibility to electrically control the performance of the device. %%
This further adds to the promise of the graphene biperiodic surfaces. %%
Based on the gained insight, useful devices can be designed for specific applications, which is the main focus of the next example. %%
In Fig.\,\ref{microwaveAbsorber}, a multilayer absorber configuration is shown, which contains five layers of graphene patches separated by SiO$_2$ thin films with thickness $t=50\,\mbox{nm}$. %%
The multilayer graphene metasurface resides on an SiO$_2$ substrate with thickness $h=3\,\mbox{mm}$ backed by a perfect electric conductor (PEC). %%
Fig.\,\ref{microwaveAbsorber}b shows the DC biasing circuit which can be used for applying the electrostatic bias field on the patches.

In Fig.\,\ref{microwaveAbsorber}c, the magnitude of the reflection coefficient in terms of frequency is depicted for various bias voltages $V_g$. %%
The structure behaves as an absorber operating at $f_0=10.7\,\mbox{GHz}$ when no bias voltage is applied. %%
When a bias voltage is applied, the perforated patch layers start to resonate, which prevents the penetration of the incident wave into the absorber. %%
Thus, a large amount of the incoming field is reflected, as depicted in Fig.\,\ref{microwaveAbsorber}c.
The large variation of reflection coefficient at resonance frequency with the bias voltage is an interesting feature of the proposed absorber, which is shown in the inset of Fig.\,\ref{microwaveAbsorber}c. %%
According to this figure, the reflected energy from the absorber is controlled by the bias voltage over a relatively large interval. %%

\subsection{Graphene polarizer}

\begin{figure}[!t]
\centering
\includegraphics[width=3.2in]{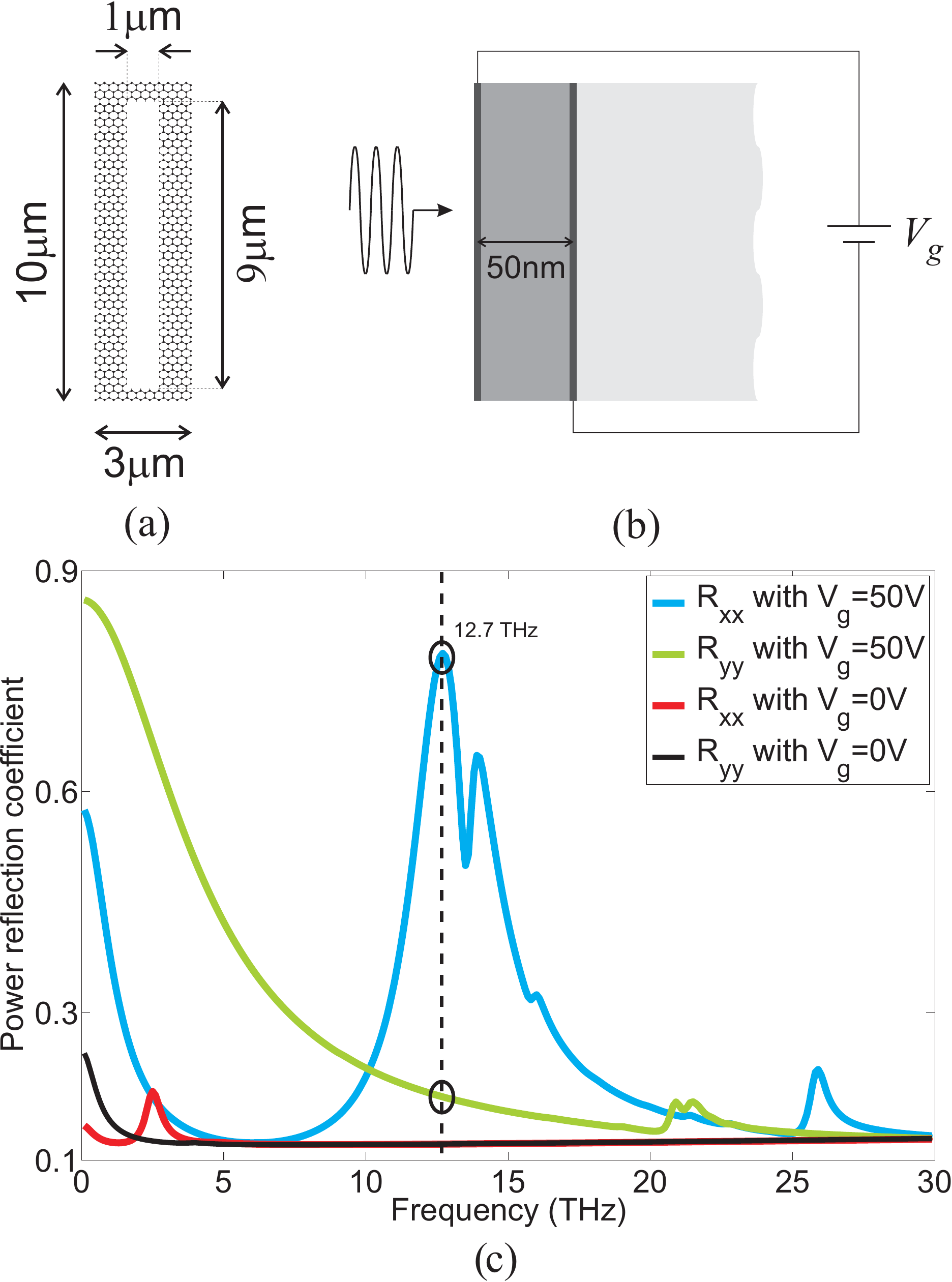}
\caption{Schematic illustration of the polarizer designed for infra-red frequencies which consists of two layers of graphene printed on a Si$_2$N$_3$ substrate: (a) The graphene patch layer, (b) the side-view of the polarizer and the biasing circuit. (c) Power reflection coefficients $R_{xx}$ and $R_{yy}$ versus frequency for bias voltages $V_g = 50\,\mbox{V}$ and $V_g = 0\,\mbox{V}$ are plotted for the depicted multilayer periodic graphene metasurface.}
\label{graphenePolarizer}
\end{figure}

A remarkable advantage of graphene biperiodic structures is that a birefringent performance can be achieved using unit cells without diagonal symmetry. %%
Based on this idea a polarizer is designed as the last example, which is illustrated in Fig.\,\ref{graphenePolarizer}. %%
Homogeneous graphene layers are etched periodically to create a lattice of slots. %%
The slots are longer along $y$ axis than $x$ axis, which leads to different performance for $x$ and $y$ polarized waves. %%
By proper adjustment of the dimensions, a tunable polarizer is designed for infrared frequencies. %%
The graphene patch layers shown in Fig.\,\ref{graphenePolarizer}a are fabricated on two sides of a Silicon Nitride ($\epsilon_r^{\mathrm{DC}}=6.6$ and $\epsilon_r^{\mathrm{IR}}=4.2$) thin film with thickness $t=50\,\mbox{nm}$ and the whole structure resides on a thick SiO$_2$ substrate (Fig.\,\ref{graphenePolarizer}b).
The reflected energy versus frequency for $V_g=0\,\mbox{V}$ and $V_g=50\,\mbox{V}$ are plotted for two polarizations in Fig.\,\ref{graphenePolarizer}c. %%
The power reflection coefficient ($R_{xx}$) represents the reflected energy for the electric $x$-polarized wave ($E_x \neq 0$, $E_y = 0$) when a similar plane wave illuminates the surface. %%
Similarly, $R_{yy}$ is computed for the electric $y$-polarized reflected and incident waves ($E_x = 0$, $E_y \neq 0$). %%
It is observed that at the resonance frequency $f=12.7\,\mbox{THz}$ the $x$ polarization is strongly reflected, whereas the $y$ polarization is weakly affected. %%
Hence, when both polarizations equally exist in the incident wave, the reflected plane wave has a dominant $x$ polarization at this frequency. %%

\section{Conclusion}

A full-vector spectral method based on the periodic method of moments (PMoM) has been introduced for the analysis of graphene biperiodic structures. %%
The method was then applied to the modeling and design of various graphene biperiodic structures, demonstrating that periodically patterned graphene structures can implement EM metasurfaces, which exhibit different and useful dynamically-controllable capabilities, influencing both the amplitude and polarization of the EM wave.

\section{Acknowledgement}

This work was supported (in part) by the Swiss National Science Foundation (SNSF) under grant n$^{\circ}$133583. The authors would also like to thank Dr. Gomez-Diaz from EPFL for the fruitful discussion.

\bibliography{../../references}

\begin{thebibliography}{47}
\expandafter\ifx\csname natexlab\endcsname\relax\def\natexlab#1{#1}\fi
\expandafter\ifx\csname bibnamefont\endcsname\relax
  \def\bibnamefont#1{#1}\fi
\expandafter\ifx\csname bibfnamefont\endcsname\relax
  \def\bibfnamefont#1{#1}\fi
\expandafter\ifx\csname citenamefont\endcsname\relax
  \def\citenamefont#1{#1}\fi
\expandafter\ifx\csname url\endcsname\relax
  \def\url#1{\texttt{#1}}\fi
\expandafter\ifx\csname urlprefix\endcsname\relax\def\urlprefix{URL }\fi
\providecommand{\bibinfo}[2]{#2}
\providecommand{\eprint}[2][]{\url{#2}}

\bibitem[{\citenamefont{Geim and Novoselov}(2007)}]{theRiseOfGraphene}
\bibinfo{author}{\bibfnamefont{A.~K.} \bibnamefont{Geim}} \bibnamefont{and}
  \bibinfo{author}{\bibfnamefont{K.~S.} \bibnamefont{Novoselov}},
  \bibinfo{journal}{Nature Mater.} \textbf{\bibinfo{volume}{6}},
  \bibinfo{pages}{183} (\bibinfo{year}{2007}).

\bibitem[{\citenamefont{Geim and Novoselov}(2009)}]{grapheneStatusAndProspects}
\bibinfo{author}{\bibfnamefont{A.~K.} \bibnamefont{Geim}} \bibnamefont{and}
  \bibinfo{author}{\bibfnamefont{K.~S.} \bibnamefont{Novoselov}},
  \bibinfo{journal}{Science} \textbf{\bibinfo{volume}{324}},
  \bibinfo{pages}{1530} (\bibinfo{year}{2009}).

\bibitem[{\citenamefont{Novoselov et~al.}(2005)\citenamefont{Novoselov, Geim,
  Morozov, Jiang, Katsnelson, Grigorieva, Dubonos, and
  Firsov}}]{masslessDiracFermions1}
\bibinfo{author}{\bibfnamefont{K.~S.} \bibnamefont{Novoselov}},
  \bibinfo{author}{\bibfnamefont{A.~K.} \bibnamefont{Geim}},
  \bibinfo{author}{\bibfnamefont{S.~V.} \bibnamefont{Morozov}},
  \bibinfo{author}{\bibfnamefont{D.}~\bibnamefont{Jiang}},
  \bibinfo{author}{\bibfnamefont{M.~I.} \bibnamefont{Katsnelson}},
  \bibinfo{author}{\bibfnamefont{I.~V.} \bibnamefont{Grigorieva}},
  \bibinfo{author}{\bibfnamefont{S.~V.} \bibnamefont{Dubonos}},
  \bibnamefont{and} \bibinfo{author}{\bibfnamefont{A.~A.}
  \bibnamefont{Firsov}}, \bibinfo{journal}{Nature}
  \textbf{\bibinfo{volume}{438}}, \bibinfo{pages}{197} (\bibinfo{year}{2005}).

\bibitem[{\citenamefont{Jiang et~al.}(2007)\citenamefont{Jiang, Henriksen,
  Tung, Wang, Schwartz, Han, Kim, and Stormer}}]{masslessDiracFermions2}
\bibinfo{author}{\bibfnamefont{Z.}~\bibnamefont{Jiang}},
  \bibinfo{author}{\bibfnamefont{E.~A.} \bibnamefont{Henriksen}},
  \bibinfo{author}{\bibfnamefont{L.~C.} \bibnamefont{Tung}},
  \bibinfo{author}{\bibfnamefont{Y.-J.} \bibnamefont{Wang}},
  \bibinfo{author}{\bibfnamefont{M.~E.} \bibnamefont{Schwartz}},
  \bibinfo{author}{\bibfnamefont{M.~Y.} \bibnamefont{Han}},
  \bibinfo{author}{\bibfnamefont{P.}~\bibnamefont{Kim}}, \bibnamefont{and}
  \bibinfo{author}{\bibfnamefont{H.~L.} \bibnamefont{Stormer}},
  \bibinfo{journal}{Phys. Rev. Lett.} \textbf{\bibinfo{volume}{98}},
  \bibinfo{pages}{197403} (\bibinfo{year}{2007}).

\bibitem[{\citenamefont{Deacon et~al.}(2007)\citenamefont{Deacon, Chuang,
  Nicholas, Novoselov, and Geim}}]{masslessDiracFermions3}
\bibinfo{author}{\bibfnamefont{R.~S.} \bibnamefont{Deacon}},
  \bibinfo{author}{\bibfnamefont{K.-C.} \bibnamefont{Chuang}},
  \bibinfo{author}{\bibfnamefont{R.~J.} \bibnamefont{Nicholas}},
  \bibinfo{author}{\bibfnamefont{K.~S.} \bibnamefont{Novoselov}},
  \bibnamefont{and} \bibinfo{author}{\bibfnamefont{A.~K.} \bibnamefont{Geim}},
  \bibinfo{journal}{Phys. Rev. B} \textbf{\bibinfo{volume}{76}},
  \bibinfo{pages}{081406} (\bibinfo{year}{2007}).

\bibitem[{\citenamefont{Li et~al.}(2008)\citenamefont{Li, Henriksen, Jiang,
  Hao, Martin, Kim, Stormer, and Basov}}]{masslessDiracFermions4}
\bibinfo{author}{\bibfnamefont{Z.}~\bibnamefont{Li}},
  \bibinfo{author}{\bibfnamefont{E.~A.} \bibnamefont{Henriksen}},
  \bibinfo{author}{\bibfnamefont{Z.}~\bibnamefont{Jiang}},
  \bibinfo{author}{\bibfnamefont{Z.}~\bibnamefont{Hao}},
  \bibinfo{author}{\bibfnamefont{M.~C.} \bibnamefont{Martin}},
  \bibinfo{author}{\bibfnamefont{P.}~\bibnamefont{Kim}},
  \bibinfo{author}{\bibfnamefont{H.~L.} \bibnamefont{Stormer}},
  \bibnamefont{and} \bibinfo{author}{\bibfnamefont{D.~N.} \bibnamefont{Basov}},
  \bibinfo{journal}{Nature Physics} \textbf{\bibinfo{volume}{4}},
  \bibinfo{pages}{532} (\bibinfo{year}{2008}).

\bibitem[{\citenamefont{Lin et~al.}(2009)\citenamefont{Lin, Jenkins,
  Valdes-Garcia, Small, Farmer, and Avouris}}]{nanoelectronicDevicesGraphene}
\bibinfo{author}{\bibfnamefont{Y.-M.} \bibnamefont{Lin}},
  \bibinfo{author}{\bibfnamefont{K.~A.} \bibnamefont{Jenkins}},
  \bibinfo{author}{\bibfnamefont{A.}~\bibnamefont{Valdes-Garcia}},
  \bibinfo{author}{\bibfnamefont{J.~P.} \bibnamefont{Small}},
  \bibinfo{author}{\bibfnamefont{D.~B.} \bibnamefont{Farmer}},
  \bibnamefont{and} \bibinfo{author}{\bibfnamefont{P.}~\bibnamefont{Avouris}},
  \bibinfo{journal}{Nano Letters} \textbf{\bibinfo{volume}{9}},
  \bibinfo{pages}{422} (\bibinfo{year}{2009}).

\bibitem[{\citenamefont{Schwierz}(2010)}]{grapheneTransistors}
\bibinfo{author}{\bibfnamefont{F.}~\bibnamefont{Schwierz}},
  \bibinfo{journal}{Nat. Nanotechnol.} \textbf{\bibinfo{volume}{5}},
  \bibinfo{pages}{487} (\bibinfo{year}{2010}).

\bibitem[{\citenamefont{Bunch et~al.}(2005)\citenamefont{Bunch, Yaish, Brink,
  Bolotin, and McEuen}}]{grapheneFabrication1}
\bibinfo{author}{\bibfnamefont{J.~S.} \bibnamefont{Bunch}},
  \bibinfo{author}{\bibfnamefont{Y.}~\bibnamefont{Yaish}},
  \bibinfo{author}{\bibfnamefont{M.}~\bibnamefont{Brink}},
  \bibinfo{author}{\bibfnamefont{K.}~\bibnamefont{Bolotin}}, \bibnamefont{and}
  \bibinfo{author}{\bibfnamefont{P.~L.} \bibnamefont{McEuen}},
  \bibinfo{journal}{Nano. Lett.} \textbf{\bibinfo{volume}{5}},
  \bibinfo{pages}{287} (\bibinfo{year}{2005}).

\bibitem[{\citenamefont{Bae et~al.}(2010)\citenamefont{Bae, Kim, Lee, Xu, Park,
  Zheng, Balakrishnan, Lei, Kim, Song et~al.}}]{grapheneFabrication2}
\bibinfo{author}{\bibfnamefont{S.}~\bibnamefont{Bae}},
  \bibinfo{author}{\bibfnamefont{H.}~\bibnamefont{Kim}},
  \bibinfo{author}{\bibfnamefont{Y.}~\bibnamefont{Lee}},
  \bibinfo{author}{\bibfnamefont{X.}~\bibnamefont{Xu}},
  \bibinfo{author}{\bibfnamefont{J.-S.} \bibnamefont{Park}},
  \bibinfo{author}{\bibfnamefont{Y.}~\bibnamefont{Zheng}},
  \bibinfo{author}{\bibfnamefont{J.}~\bibnamefont{Balakrishnan}},
  \bibinfo{author}{\bibfnamefont{T.}~\bibnamefont{Lei}},
  \bibinfo{author}{\bibfnamefont{H.~R.} \bibnamefont{Kim}},
  \bibinfo{author}{\bibfnamefont{Y.~I.} \bibnamefont{Song}},
  \bibnamefont{et~al.}, \bibinfo{journal}{Nat. Nanotechnol.}
  \textbf{\bibinfo{volume}{5}}, \bibinfo{pages}{574} (\bibinfo{year}{2010}).

\bibitem[{\citenamefont{Gomez-Diaz et~al.}(2012)\citenamefont{Gomez-Diaz,
  Perruisseau-Carrier, Sharma, and Ionescu}}]{grapheneFabrication3}
\bibinfo{author}{\bibfnamefont{J.}~\bibnamefont{Gomez-Diaz}},
  \bibinfo{author}{\bibfnamefont{J.}~\bibnamefont{Perruisseau-Carrier}},
  \bibinfo{author}{\bibfnamefont{P.}~\bibnamefont{Sharma}}, \bibnamefont{and}
  \bibinfo{author}{\bibfnamefont{A.~M.} \bibnamefont{Ionescu}},
  \bibinfo{journal}{J. of Appl. Phys.} \textbf{\bibinfo{volume}{111}},
  \bibinfo{pages}{114908} (\bibinfo{year}{2012}).

\bibitem[{\citenamefont{Vakil and Engheta}(2011)}]{grapheneMetamaterials1}
\bibinfo{author}{\bibfnamefont{A.}~\bibnamefont{Vakil}} \bibnamefont{and}
  \bibinfo{author}{\bibfnamefont{N.}~\bibnamefont{Engheta}},
  \bibinfo{journal}{Science} \textbf{\bibinfo{volume}{332}},
  \bibinfo{pages}{1291} (\bibinfo{year}{2011}).

\bibitem[{\citenamefont{Papasimakis et~al.}(2010)\citenamefont{Papasimakis,
  Luo, Shen, Angelis, Fabrizio, Nikolaenko, and
  Zheludev}}]{grapheneMetamaterials2}
\bibinfo{author}{\bibfnamefont{N.}~\bibnamefont{Papasimakis}},
  \bibinfo{author}{\bibfnamefont{Z.}~\bibnamefont{Luo}},
  \bibinfo{author}{\bibfnamefont{Z.~X.} \bibnamefont{Shen}},
  \bibinfo{author}{\bibfnamefont{F.~D.} \bibnamefont{Angelis}},
  \bibinfo{author}{\bibfnamefont{E.~D.} \bibnamefont{Fabrizio}},
  \bibinfo{author}{\bibfnamefont{A.~E.} \bibnamefont{Nikolaenko}},
  \bibnamefont{and} \bibinfo{author}{\bibfnamefont{N.~I.}
  \bibnamefont{Zheludev}}, \bibinfo{journal}{Opt. Express}
  \textbf{\bibinfo{volume}{18}}, \bibinfo{pages}{8353} (\bibinfo{year}{2010}).

\bibitem[{\citenamefont{Huang et~al.}(2011)\citenamefont{Huang, Wu, and
  Mao}}]{grapheneMetamaterials3}
\bibinfo{author}{\bibfnamefont{Y.}~\bibnamefont{Huang}},
  \bibinfo{author}{\bibfnamefont{L.-S.} \bibnamefont{Wu}}, \bibnamefont{and}
  \bibinfo{author}{\bibfnamefont{J.}~\bibnamefont{Mao}}, in
  \emph{\bibinfo{booktitle}{Electromagnetics in Advanced Applications (ICEAA),
  2011 International Conference on}} (\bibinfo{year}{2011}), pp.
  \bibinfo{pages}{556 --559}.

\bibitem[{\citenamefont{Chen and Al{\'u}}(2011)}]{grapheneMetamaterials4}
\bibinfo{author}{\bibfnamefont{P.-Y.} \bibnamefont{Chen}} \bibnamefont{and}
  \bibinfo{author}{\bibfnamefont{A.}~\bibnamefont{Al{\'u}}},
  \bibinfo{journal}{ACS Nano} \textbf{\bibinfo{volume}{5}},
  \bibinfo{pages}{5855} (\bibinfo{year}{2011}).

\bibitem[{\citenamefont{Gomez-Diaz and
  Perruisseau-Carrier}(2012)}]{sebasProceedingPaper}
\bibinfo{author}{\bibfnamefont{J.~S.} \bibnamefont{Gomez-Diaz}}
  \bibnamefont{and}
  \bibinfo{author}{\bibfnamefont{J.}~\bibnamefont{Perruisseau-Carrier}}, in
  \emph{\bibinfo{booktitle}{to appear in Proc. ISAP2012}}
  (\bibinfo{address}{Nagoya, Japan}, \bibinfo{year}{2012}).

\bibitem[{\citenamefont{Christensen et~al.}(2012)\citenamefont{Christensen,
  Manjavacas, Thongrattanasiri, Koppens, and de~Abajo}}]{grapheneStripes1}
\bibinfo{author}{\bibfnamefont{J.}~\bibnamefont{Christensen}},
  \bibinfo{author}{\bibfnamefont{A.}~\bibnamefont{Manjavacas}},
  \bibinfo{author}{\bibfnamefont{S.}~\bibnamefont{Thongrattanasiri}},
  \bibinfo{author}{\bibfnamefont{F.~H.~L.} \bibnamefont{Koppens}},
  \bibnamefont{and} \bibinfo{author}{\bibfnamefont{F.~J.~G.}
  \bibnamefont{de~Abajo}}, \bibinfo{journal}{ACS Nano}
  \textbf{\bibinfo{volume}{6}}, \bibinfo{pages}{431} (\bibinfo{year}{2012}).

\bibitem[{\citenamefont{Ju et~al.}(2011)\citenamefont{Ju, Geng, Horng, Girit,
  Martin, Hao, Bechtel, Liang, Zettl, Shen et~al.}}]{grapheneStripes2}
\bibinfo{author}{\bibfnamefont{L.}~\bibnamefont{Ju}},
  \bibinfo{author}{\bibfnamefont{B.}~\bibnamefont{Geng}},
  \bibinfo{author}{\bibfnamefont{J.}~\bibnamefont{Horng}},
  \bibinfo{author}{\bibfnamefont{C.}~\bibnamefont{Girit}},
  \bibinfo{author}{\bibfnamefont{M.}~\bibnamefont{Martin}},
  \bibinfo{author}{\bibfnamefont{Z.}~\bibnamefont{Hao}},
  \bibinfo{author}{\bibfnamefont{H.~A.} \bibnamefont{Bechtel}},
  \bibinfo{author}{\bibfnamefont{X.}~\bibnamefont{Liang}},
  \bibinfo{author}{\bibfnamefont{A.}~\bibnamefont{Zettl}},
  \bibinfo{author}{\bibfnamefont{Y.~R.} \bibnamefont{Shen}},
  \bibnamefont{et~al.}, \bibinfo{journal}{Nat. Nanotechnol.}
  \textbf{\bibinfo{volume}{6}}, \bibinfo{pages}{630} (\bibinfo{year}{2011}).

\bibitem[{\citenamefont{Yan et~al.}(2012)\citenamefont{Yan, Li, Chandra,
  Tulevski, Wu, Freitag, Zhu, Avouris, and Xia}}]{patternedGrapheneNature}
\bibinfo{author}{\bibfnamefont{H.}~\bibnamefont{Yan}},
  \bibinfo{author}{\bibfnamefont{X.}~\bibnamefont{Li}},
  \bibinfo{author}{\bibfnamefont{B.}~\bibnamefont{Chandra}},
  \bibinfo{author}{\bibfnamefont{G.}~\bibnamefont{Tulevski}},
  \bibinfo{author}{\bibfnamefont{Y.}~\bibnamefont{Wu}},
  \bibinfo{author}{\bibfnamefont{M.}~\bibnamefont{Freitag}},
  \bibinfo{author}{\bibfnamefont{W.}~\bibnamefont{Zhu}},
  \bibinfo{author}{\bibfnamefont{P.}~\bibnamefont{Avouris}}, \bibnamefont{and}
  \bibinfo{author}{\bibfnamefont{F.}~\bibnamefont{Xia}}, \bibinfo{journal}{Nat.
  Nanotechnol.} \textbf{\bibinfo{volume}{7}}, \bibinfo{pages}{330}
  (\bibinfo{year}{2012}).

\bibitem[{\citenamefont{Thongrattanasiri
  et~al.}(2012)\citenamefont{Thongrattanasiri, Koppens, and
  de~Abajo}}]{grapheneAbsorption1}
\bibinfo{author}{\bibfnamefont{S.}~\bibnamefont{Thongrattanasiri}},
  \bibinfo{author}{\bibfnamefont{F.~H.~L.} \bibnamefont{Koppens}},
  \bibnamefont{and} \bibinfo{author}{\bibfnamefont{F.~J.~G.}
  \bibnamefont{de~Abajo}}, \bibinfo{journal}{Phys. Rev. Lett.}
  \textbf{\bibinfo{volume}{108}}, \bibinfo{pages}{047401}
  (\bibinfo{year}{2012}).

\bibitem[{\citenamefont{Nikitin
  et~al.}(2012{\natexlab{a}})\citenamefont{Nikitin, Guinea, Garcia-Vidal, and
  Martin-Moreno}}]{grapheneAbsorption2}
\bibinfo{author}{\bibfnamefont{A.~Y.} \bibnamefont{Nikitin}},
  \bibinfo{author}{\bibfnamefont{F.}~\bibnamefont{Guinea}},
  \bibinfo{author}{\bibfnamefont{F.~J.} \bibnamefont{Garcia-Vidal}},
  \bibnamefont{and}
  \bibinfo{author}{\bibfnamefont{L.}~\bibnamefont{Martin-Moreno}},
  \bibinfo{journal}{Phys. Rev. B} \textbf{\bibinfo{volume}{85}},
  \bibinfo{pages}{081405} (\bibinfo{year}{2012}{\natexlab{a}}).

\bibitem[{\citenamefont{Nikitin
  et~al.}(2012{\natexlab{b}})\citenamefont{Nikitin, Guinea, and
  Martin-Moreno}}]{grapheneAbsorption3}
\bibinfo{author}{\bibfnamefont{A.~Y.} \bibnamefont{Nikitin}},
  \bibinfo{author}{\bibfnamefont{F.}~\bibnamefont{Guinea}}, \bibnamefont{and}
  \bibinfo{author}{\bibfnamefont{L.}~\bibnamefont{Martin-Moreno}}
  (\bibinfo{year}{2012}{\natexlab{b}}), \eprint{arXiv:1206.2163v1}.

\bibitem[{\citenamefont{Ferreira and Peres}(2012)}]{grapheneAbsorption4}
\bibinfo{author}{\bibfnamefont{A.}~\bibnamefont{Ferreira}} \bibnamefont{and}
  \bibinfo{author}{\bibfnamefont{N.~M.~R.} \bibnamefont{Peres}}
  (\bibinfo{year}{2012}), \eprint{arXiv:1206.3854v1}.

\bibitem[{\citenamefont{Wu}(1995)}]{bookFSS1}
\bibinfo{editor}{\bibfnamefont{T.~K.} \bibnamefont{Wu}}, ed.,
  \emph{\bibinfo{title}{Frequency Selective Surface and Grid Array}}
  (\bibinfo{publisher}{John Wiley and Sons}, \bibinfo{address}{New York, NY},
  \bibinfo{year}{1995}).

\bibitem[{\citenamefont{Munk}(2000)}]{bookFSS2}
\bibinfo{author}{\bibfnamefont{B.~A.} \bibnamefont{Munk}},
  \emph{\bibinfo{title}{Frequency Selective Surfaces Theory and Design}}
  (\bibinfo{publisher}{John Wiley and Sons}, \bibinfo{address}{New York, NY},
  \bibinfo{year}{2000}).

\bibitem[{\citenamefont{Hu et~al.}(2007)\citenamefont{Hu, Dickie, Cahill,
  Gamble, Ismail, Fusco, Linton, Grant, and Rea}}]{dynamicFSS1}
\bibinfo{author}{\bibfnamefont{W.}~\bibnamefont{Hu}},
  \bibinfo{author}{\bibfnamefont{R.}~\bibnamefont{Dickie}},
  \bibinfo{author}{\bibfnamefont{R.}~\bibnamefont{Cahill}},
  \bibinfo{author}{\bibfnamefont{H.}~\bibnamefont{Gamble}},
  \bibinfo{author}{\bibfnamefont{Y.}~\bibnamefont{Ismail}},
  \bibinfo{author}{\bibfnamefont{V.}~\bibnamefont{Fusco}},
  \bibinfo{author}{\bibfnamefont{D.}~\bibnamefont{Linton}},
  \bibinfo{author}{\bibfnamefont{N.}~\bibnamefont{Grant}}, \bibnamefont{and}
  \bibinfo{author}{\bibfnamefont{S.}~\bibnamefont{Rea}}, \bibinfo{journal}{IEEE
  Microw. Wirel. Co.} \textbf{\bibinfo{volume}{17}}, \bibinfo{pages}{667 }
  (\bibinfo{year}{2007}).

\bibitem[{\citenamefont{Mias}(2005)}]{dynamicFSS2}
\bibinfo{author}{\bibfnamefont{C.}~\bibnamefont{Mias}}, \bibinfo{journal}{IEEE
  Microw. Wirel. Co.} \textbf{\bibinfo{volume}{15}}, \bibinfo{pages}{570 }
  (\bibinfo{year}{2005}).

\bibitem[{\citenamefont{Schoenlinner et~al.}(2004)\citenamefont{Schoenlinner,
  Abbaspour-Tamijani, Kempel, and Rebeiz}}]{dynamicFSS3}
\bibinfo{author}{\bibfnamefont{B.}~\bibnamefont{Schoenlinner}},
  \bibinfo{author}{\bibfnamefont{A.}~\bibnamefont{Abbaspour-Tamijani}},
  \bibinfo{author}{\bibfnamefont{L.~C.} \bibnamefont{Kempel}},
  \bibnamefont{and} \bibinfo{author}{\bibfnamefont{G.~M.}
  \bibnamefont{Rebeiz}}, \bibinfo{journal}{IEEE T. Microw. Theory}
  \textbf{\bibinfo{volume}{52}}, \bibinfo{pages}{2474} (\bibinfo{year}{2004}).

\bibitem[{\citenamefont{Lei et~al.}(2011)\citenamefont{Lei, Zamora, Chun, Ohta,
  and Shiroma}}]{dynamicFSS4}
\bibinfo{author}{\bibfnamefont{B.~J.} \bibnamefont{Lei}},
  \bibinfo{author}{\bibfnamefont{A.}~\bibnamefont{Zamora}},
  \bibinfo{author}{\bibfnamefont{T.}~\bibnamefont{Chun}},
  \bibinfo{author}{\bibfnamefont{A.}~\bibnamefont{Ohta}}, \bibnamefont{and}
  \bibinfo{author}{\bibfnamefont{W.}~\bibnamefont{Shiroma}},
  \bibinfo{journal}{IEEE Microw. Wirel. Co.} \textbf{\bibinfo{volume}{21}},
  \bibinfo{pages}{465 } (\bibinfo{year}{2011}).

\bibitem[{\citenamefont{Gusynin et~al.}(2007)\citenamefont{Gusynin, Sharapov,
  and Carbotte}}]{grapheneConductivityModel}
\bibinfo{author}{\bibfnamefont{V.~P.} \bibnamefont{Gusynin}},
  \bibinfo{author}{\bibfnamefont{S.~G.} \bibnamefont{Sharapov}},
  \bibnamefont{and} \bibinfo{author}{\bibfnamefont{J.~P.}
  \bibnamefont{Carbotte}}, \bibinfo{journal}{J. Phys.-Condens. Mat.}
  \textbf{\bibinfo{volume}{19}}, \bibinfo{pages}{026222}
  (\bibinfo{year}{2007}).

\bibitem[{\citenamefont{Falkovsky}(2008)}]{grapheneOpticalProperties}
\bibinfo{author}{\bibfnamefont{L.~A.} \bibnamefont{Falkovsky}},
  \bibinfo{journal}{J. Phys. Conf. Ser.} \textbf{\bibinfo{volume}{129}},
  \bibinfo{pages}{012004} (\bibinfo{year}{2008}).

\bibitem[{\citenamefont{Sensale-Rodriguez
  et~al.}(2012)\citenamefont{Sensale-Rodriguez, Yan, Kelly, Fang, Tahy, Hwang,
  Jena, Liu, and Xing}}]{grapheneModulators}
\bibinfo{author}{\bibfnamefont{B.}~\bibnamefont{Sensale-Rodriguez}},
  \bibinfo{author}{\bibfnamefont{R.}~\bibnamefont{Yan}},
  \bibinfo{author}{\bibfnamefont{M.~M.} \bibnamefont{Kelly}},
  \bibinfo{author}{\bibfnamefont{T.}~\bibnamefont{Fang}},
  \bibinfo{author}{\bibfnamefont{K.}~\bibnamefont{Tahy}},
  \bibinfo{author}{\bibfnamefont{W.~S.} \bibnamefont{Hwang}},
  \bibinfo{author}{\bibfnamefont{D.}~\bibnamefont{Jena}},
  \bibinfo{author}{\bibfnamefont{L.}~\bibnamefont{Liu}}, \bibnamefont{and}
  \bibinfo{author}{\bibfnamefont{H.~G.} \bibnamefont{Xing}},
  \bibinfo{journal}{Nature Comm.} \textbf{\bibinfo{volume}{3}}
  (\bibinfo{year}{2012}).

\bibitem[{\citenamefont{Hanson}(2008{\natexlab{a}})}]{grapheneModeling3}
\bibinfo{author}{\bibfnamefont{G.}~\bibnamefont{Hanson}},
  \bibinfo{journal}{IEEE T. Antenn. Propag.} \textbf{\bibinfo{volume}{56}},
  \bibinfo{pages}{747 } (\bibinfo{year}{2008}{\natexlab{a}}).

\bibitem[{\citenamefont{Slepyan et~al.}(1999)\citenamefont{Slepyan, Maksimenko,
  Lakhtakia, Yevtushenko, and Gusakov}}]{carbonNanotubes}
\bibinfo{author}{\bibfnamefont{G.~Y.} \bibnamefont{Slepyan}},
  \bibinfo{author}{\bibfnamefont{S.~A.} \bibnamefont{Maksimenko}},
  \bibinfo{author}{\bibfnamefont{A.}~\bibnamefont{Lakhtakia}},
  \bibinfo{author}{\bibfnamefont{O.}~\bibnamefont{Yevtushenko}},
  \bibnamefont{and} \bibinfo{author}{\bibfnamefont{A.~V.}
  \bibnamefont{Gusakov}}, \bibinfo{journal}{Phys. Rev. B}
  \textbf{\bibinfo{volume}{60}}, \bibinfo{pages}{17136} (\bibinfo{year}{1999}).

\bibitem[{\citenamefont{Falkovsky and
  Pershoguba}(2007)}]{grapheneTheoreticalModeling1}
\bibinfo{author}{\bibfnamefont{L.~A.} \bibnamefont{Falkovsky}}
  \bibnamefont{and} \bibinfo{author}{\bibfnamefont{S.~S.}
  \bibnamefont{Pershoguba}}, \bibinfo{journal}{Phys. Rev. B}
  \textbf{\bibinfo{volume}{76}}, \bibinfo{pages}{153410}
  (\bibinfo{year}{2007}).

\bibitem[{\citenamefont{Gusynin and
  Sharapov}(2006)}]{grapheneTheoreticalModeling2}
\bibinfo{author}{\bibfnamefont{V.~P.} \bibnamefont{Gusynin}} \bibnamefont{and}
  \bibinfo{author}{\bibfnamefont{S.~G.} \bibnamefont{Sharapov}},
  \bibinfo{journal}{Phys. Rev. B} \textbf{\bibinfo{volume}{73}},
  \bibinfo{pages}{245411} (\bibinfo{year}{2006}).

\bibitem[{\citenamefont{Gusynin et~al.}(2006)\citenamefont{Gusynin, Sharapov,
  and Carbotte}}]{grapheneTheoreticalModeling3}
\bibinfo{author}{\bibfnamefont{V.~P.} \bibnamefont{Gusynin}},
  \bibinfo{author}{\bibfnamefont{S.~G.} \bibnamefont{Sharapov}},
  \bibnamefont{and} \bibinfo{author}{\bibfnamefont{J.~P.}
  \bibnamefont{Carbotte}}, \bibinfo{journal}{Phys. Rev. Lett.}
  \textbf{\bibinfo{volume}{96}}, \bibinfo{pages}{256802}
  (\bibinfo{year}{2006}).

\bibitem[{\citenamefont{Ziegler}(2007)}]{grapheneTheoreticalModeling4}
\bibinfo{author}{\bibfnamefont{K.}~\bibnamefont{Ziegler}},
  \bibinfo{journal}{Phys. Rev. B} \textbf{\bibinfo{volume}{75}},
  \bibinfo{pages}{233407} (\bibinfo{year}{2007}).

\bibitem[{\citenamefont{Kim et~al.}(2011)\citenamefont{Kim, Lee, Bae, Kim,
  Hong, and Choi}}]{grapheneSubstrateEffect}
\bibinfo{author}{\bibfnamefont{J.~Y.} \bibnamefont{Kim}},
  \bibinfo{author}{\bibfnamefont{C.}~\bibnamefont{Lee}},
  \bibinfo{author}{\bibfnamefont{S.}~\bibnamefont{Bae}},
  \bibinfo{author}{\bibfnamefont{K.~S.} \bibnamefont{Kim}},
  \bibinfo{author}{\bibfnamefont{B.~H.} \bibnamefont{Hong}}, \bibnamefont{and}
  \bibinfo{author}{\bibfnamefont{E.~J.} \bibnamefont{Choi}},
  \bibinfo{journal}{Appl. Phys. Lett.} \textbf{\bibinfo{volume}{98}},
  \bibinfo{pages}{201907} (\bibinfo{year}{2011}).

\bibitem[{\citenamefont{Hanson}(2008{\natexlab{b}})}]{grapheneModeling2}
\bibinfo{author}{\bibfnamefont{G.~W.} \bibnamefont{Hanson}},
  \bibinfo{journal}{J. of Appl. Phys.} \textbf{\bibinfo{volume}{103}},
  \bibinfo{pages}{064302 } (\bibinfo{year}{2008}{\natexlab{b}}).

\bibitem[{\citenamefont{Lee et~al.}(2011)\citenamefont{Lee, Kim, Bae, Kim,
  Hong, and Choi}}]{GammaValue}
\bibinfo{author}{\bibfnamefont{C.}~\bibnamefont{Lee}},
  \bibinfo{author}{\bibfnamefont{J.~Y.} \bibnamefont{Kim}},
  \bibinfo{author}{\bibfnamefont{S.}~\bibnamefont{Bae}},
  \bibinfo{author}{\bibfnamefont{K.~S.} \bibnamefont{Kim}},
  \bibinfo{author}{\bibfnamefont{B.~H.} \bibnamefont{Hong}}, \bibnamefont{and}
  \bibinfo{author}{\bibfnamefont{E.~J.} \bibnamefont{Choi}},
  \bibinfo{journal}{Appl. Phys. Lett.} \textbf{\bibinfo{volume}{98}},
  \bibinfo{pages}{071905} (\bibinfo{year}{2011}).

\bibitem[{\citenamefont{Hanson et~al.}(2011)\citenamefont{Hanson, Yakovlev, and
  Mafi}}]{grapheneModeling1}
\bibinfo{author}{\bibfnamefont{G.~W.} \bibnamefont{Hanson}},
  \bibinfo{author}{\bibfnamefont{A.~B.} \bibnamefont{Yakovlev}},
  \bibnamefont{and} \bibinfo{author}{\bibfnamefont{A.}~\bibnamefont{Mafi}},
  \bibinfo{journal}{J. of Appl. Phys.} \textbf{\bibinfo{volume}{110}},
  \bibinfo{pages}{114305 } (\bibinfo{year}{2011}).

\bibitem[{\citenamefont{Lovat}(2012)}]{grapheneModeling4}
\bibinfo{author}{\bibfnamefont{G.}~\bibnamefont{Lovat}}, \bibinfo{journal}{IEEE
  T. Electromagn. C.} \textbf{\bibinfo{volume}{54}}, \bibinfo{pages}{101 }
  (\bibinfo{year}{2012}).

\bibitem[{\citenamefont{Fallahi}(1964)}]{myThesis}
\bibinfo{author}{\bibfnamefont{A.}~\bibnamefont{Fallahi}}, Ph.D. thesis,
  \bibinfo{school}{ETH Z\"urich}, \bibinfo{address}{Z\"urich, Switzerland}
  (\bibinfo{year}{1964}).

\bibitem[{\citenamefont{Lin et~al.}(2006)\citenamefont{Lin, Liu, and
  Yuan}}]{anisoFSSanalysis1}
\bibinfo{author}{\bibfnamefont{B.}~\bibnamefont{Lin}},
  \bibinfo{author}{\bibfnamefont{S.}~\bibnamefont{Liu}}, \bibnamefont{and}
  \bibinfo{author}{\bibfnamefont{N.}~\bibnamefont{Yuan}},
  \bibinfo{journal}{IEEE T. Antenn. Propag.} \textbf{\bibinfo{volume}{54}},
  \bibinfo{pages}{674} (\bibinfo{year}{2006}).

\bibitem[{\citenamefont{Fallahi et~al.}(2009)\citenamefont{Fallahi, Mishrikey,
  Hafner, and Vahldieck}}]{anisoPeriodicFSSAnalysis}
\bibinfo{author}{\bibfnamefont{A.}~\bibnamefont{Fallahi}},
  \bibinfo{author}{\bibfnamefont{M.}~\bibnamefont{Mishrikey}},
  \bibinfo{author}{\bibfnamefont{C.}~\bibnamefont{Hafner}}, \bibnamefont{and}
  \bibinfo{author}{\bibfnamefont{R.}~\bibnamefont{Vahldieck}},
  \bibinfo{journal}{Metamaterials} \textbf{\bibinfo{volume}{3}},
  \bibinfo{pages}{63 } (\bibinfo{year}{2009}).

\bibitem[{\citenamefont{Johnson et~al.}(1969)\citenamefont{Johnson, Sherman,
  and Weil}}]{GaAsOpticalMeasurement}
\bibinfo{author}{\bibfnamefont{C.~J.} \bibnamefont{Johnson}},
  \bibinfo{author}{\bibfnamefont{G.~H.} \bibnamefont{Sherman}},
  \bibnamefont{and} \bibinfo{author}{\bibfnamefont{R.}~\bibnamefont{Weil}},
  \bibinfo{journal}{Appl. Opt.} \textbf{\bibinfo{volume}{8}},
  \bibinfo{pages}{1667} (\bibinfo{year}{1969}).

\end{thebibliography}

\end{document}